\documentclass[11pt,eqsecnum]{article}
\usepackage{amsmath,amsfonts,amssymb,color,graphicx,latexsym,theorem,mathrsfs,latexsym,cite,setspace
}

\newcommand{\ma}[1]{\mbox{$\mathcal{#1}$}}
\newcommand{\mas}[1]{\mbox{$\mathscr{#1}$}}

\newcommand{\D}{{\rm d}}
\newcommand{\ti}{\tilde}
\newcommand{\we}{\wedge}

\begin{document}

\begin{titlepage}

\begin{flushright}
{
\small IFUM-1044-FT\\
KOBE-TH-15-11\\
\today
}
\end{flushright}
\vspace{1cm}

\begin{center}
{\LARGE \bf
\begin{spacing}{1}
Killing-Yano tensor and supersymmetry of the self-dual Plebanski-Demianski solution
\end{spacing}
}
\end{center}
\vspace{.5cm}

\begin{center}
{\large 
Masato Nozawa$^{1,2}$ and 
Tsuyoshi Houri$^3$
} \\

\vspace{.4cm}

{\it ${}^1$Dipartimento di Fisica, Universit\`a di Milano,
and INFN, Sezione di Milano, \\
Via Celoria 16, 20133 Milano, Italia \\ 
${}^2$Department of Physics, Kyoto University, Kyoto, 606-8502, Japan\\
${}^3$Department of Physics, Kobe University, Kobe, 657-8501 Japan \\
}

\end{center}

\vspace{.5cm}

\begin{abstract}
We explore various aspects of the self-dual Pleba\'nski-Demia\'nski family in the Euclidean Einstein-Maxwell-$\Lambda$ system. The Killing-Yano tensor which was recently found by Yasui and one of the present authors allows us to prove that the self-dual Pleba\'nski-Demia\'nski metric can be brought into the self-dual Carter metric by an orientation-reversing coordinate transformation. We show that the self-dual Pleba\'nski-Demia\'nski solution admits two independent Killing spinors in the framework of $N=2$ minimal gauged supergravity, whereas the non-self-dual solution admits only a single Killing spinor. This can be demonstrated by casting the self-dual Pleba\'nski-Demia\'nski metric into two distinct Przanowski-Tod forms. As a by-product, a new example of the three-dimensional Einstein-Weyl space is presented. We also prove that the self-dual Pleba\'nski-Demia\'nski metric falls into two different Calderbank-Pedersen families, which are determined by a single function subjected to a linear equation on the two dimensional hyperbolic space. Furthermore, we consider the hyper-K\"ahler case for which the metric falls into the Gibbons-Hawking class.  We find that the condition for the nonexistence of Dirac-Misner string enforces the solution with a nonvanishing acceleration parameter to the Eguchi-Hanson space. 
\end{abstract}

\vspace{.5cm}

\setcounter{footnote}{0}

\end{titlepage}

\tableofcontents

\section{Introduction}
The concept of hidden symmetry has played a distinguished role in many problems of physics. Of most prominence in general relativity is the Killing tensor~\cite{Walker:1970un} living in the geometry of a Kerr black hole. This rank-two tensorial field assures the integrability of geodesic motion and scalar wave equation~\cite{Carter:1968ks,Hughston:1972qf,Carter:1977pq}, which cannot be captured only by Killing vectors. Various problems around black holes are  amendable to analytic study thanks to the separation constant, referred to as a Carter constant, generated by the Killing tensor. Furthermore, Floyd~\cite{Floyd} and Penrose~\cite{Penrose:1973um} discovered that the Killing tensor of the Kerr(-Newman) metric consists of the square of the anti-symmetric Killing-Yano (KY) tensor~\cite{Yano}, which underlies the integrability of Dirac equation~\cite{Carter:1979fe}. Over the last decade, these issues have been generalized to higher dimensions and intensively studied by many authors (cf. \cite{Frolov:2008jr,Yasui:2011pr,Cariglia:2014ysa} for a comprehensive review). 

From a variety of points of view, the KY tensor is more fundamental than a Killing tensor. For instance, the KY tensor generates supersymmetry for the motion of a spinning particle in curved space~\cite{Gibbons:1993ap}. At the quantum level, KY tensors correspond to the symmetry generators without anomaly~\cite{Carter:1979fe,Benn:1996ia,Benn2004,Cariglia:2003kf} unlike the Killing tensors. Papadopoulos analyzed the KY tensors in the context of manifolds with $G\subset{\rm Spin}(n)$ structures~\cite{Papadopoulos:2007gf}. Transformations of KY tensor under T-duality and its relation to the double field theory were demystified in \cite{Chervonyi:2015ima}. 
Moreover, the existence of a nondegenerate rank-two closed conformal Killing-Yano (CKY) tensor, which is dual to the rank $d-2$ KY tensor in $d$ dimensions, determines completely the local metric~\cite{Houri:2007xz,Houri:2012eq} which allows the separation of variables for Hamilton-Jacobi equations \cite{Benenti}, and fixes the algebraic type of curvatures.  

Recently, Yasui and one of the present authors investigated the spacetime (hidden) symmetries by utilizing the idea of curvature conditions~\cite{Houri:2014hma}. They focused upon the Killing connections on the vector bundle, whose parallel sections one-to-one correspond to form-fields generating the symmetry. The investigation of the curvature conditions for this vector bundle enables us to obtain the maximal number of (conformal) Killing vectors and (conformal) KY tensors. Although this scheme merely gives rise to the necessary conditions for the existence of symmetry generators, this method turns out very powerful since the curvature conditions constrain the possible form of symmetries tightly. In their paper \cite{Houri:2014hma}, KY tensors of the Pleba\'nski-Demia\'nski (PD) family~\cite{PD} were addressed. The PD metric in Lorentzian signature is the most general Petrov-D electrovacuum spacetime with an aligned non-null Maxwell field in four dimensions~\cite{Kinnersley:1969zza,DKM}, and describes a rotating, uniformly accelerating charged point source~\cite{Griffiths:2005qp}. It has been widely believed that the general PD metric with a nonvanishing acceleration parameter fails to admit any rank-two Killing tensors besides the metric, since the geodesic motions were unable to separate.\footnote{Since the PD metric belongs to Petrov-D class, it has been known that it admits a conformal Killing tensor constructed out of a CKY tensor~\cite{Walker:1970un,Kubiznak:2007kh}. The CKY tensor in the PD spacetime allows the integrability and the separation of variables only for conformally invariant field equations. See e.g \cite{Nozawa:2008wf}.} Nevertheless, when the Weyl curvature of the PD metric is self-dual, ref. \cite{Houri:2014hma} found a nondegenerate rank-two KY tensor in the self-dual PD metric. This comes out as a surprise since according to the theorem given in ref.~\cite{Houri:2007xz}, a four-dimensional manifold admitting a nondegenerate rank-two KY tensor must fall into the Carter family, which is the zero acceleration limit of the PD family. The profound mathematical reason for the existence of a nondegenerate KY tensor remains mysterious and unresolved in~\cite{Houri:2007xz}. To fill this gap is one of the aims of the present paper. 

The notion of selfdual curvatures is absent for Lorentzian signature.  We have several physical motivations for studying these geometries. A prime example is the gravitational instantons admitting a self-dual Riemann tensor~\cite{Atiyah:1978wi}. These solutions are reminiscent of self-dual instantons in Yang-Mills theory and play some essential roles in quantum gravity. Under the saddle point approximations for the path integral formulation, quantum amplitudes defined by all sums over positive-definite metrics are dominated by contributions from gravitational instantons~\cite{Eguchi:1980jx}. Mathematically, four-dimensional Euclidean spaces with a self-dual Riemann tensor are referred to as the hyper-K\"ahler manifolds.  Manifolds with a self-dual Weyl tensor are the natural generalization of instantons for non Ricci-flat spaces and share some of their properties~\cite{Gibbons:1978zy}. Another impetus for examining these geometries comes from the context of supersymmetry. Spaces with self-dual curvatures have a reduced holonomy and are instrumental for possible internal spaces of supergravity solutions. Einstein manifolds with a self-dual Weyl tensor arise as supersymmetric solutions in Euclidean supergravity~\cite{Dunajski:2010zp,Gutowski:2010zs,Dunajski:2010uv}, which have been an active subject from the standpoint of the localization principle~\cite{Martelli:2012sz,Martelli:2013aqa,Farquet:2014kma,Hama:2011ea,Hosomichi:2015jta}.  In addition, Einstein manifolds with a self-dual Weyl tensor admit a quaternionic structure, which is related to twistor spaces~\cite{Hoegner:2012sq} and arises as a nonlinear sigma model for hypermultiplets in $N=2$ supergravity~\cite{FVP}.  The superconformal calculus uncovered that quaternion-K\"ahler spaces are associated with the hyper-K\"ahler cones by the $N=2$ superconformal quotient~\cite{deWit:2001dj}. Furthermore, any Riemannian manifold with a self-dual Weyl tensor serves as a solution to the equations of motion of the conformal gravity given by the squared norm of the Weyl tensor \cite{Dunajski:2014}.

This article is intended to deepen our understanding of the hidden symmetry, supersymmetry and gravitational instantons in four-dimensional Einstein-Maxwell theory with a cosmological constant. In particular, we concentrate on the self-dual PD metric. We first notice that the general PD metric is conformally toric K\"ahler. Using the CKY tensor equipped with the Einstein manifold of self-dual Weyl tensor, it is shown that the KY tensor of PD metric found in \cite{Houri:2014hma} emerges naturally together with the type-D characteristics. Making use of the eigenvalues of the KY tensor, we identify the explicit coordinate transformation, which brings the self-dual PD family into the self-dual Carter family with opposite orientation. Namely, there exist two ways of writing the most general Petrov-D Einstein metric with a self-dual Weyl curvature in the forms of the PD metric and the Carter metric. The sign flip of orientation traces back to the ambi-K\"ahler structure. It turns out that this is closely related to the fact that the solution keeps half of supersymmetry, while the non-self-dual solution preserves only one quarter of supersymmetry. We construct explicit Killing spinors by translating the self-dual metric into the Przanowski-Tod form~\cite{Przanowski:1991ru,Tod:2006wj} in two different manners. This gives a new instance of three-dimensional Einstein-Weyl space. We are further able to rewrite the self-dual PD metric into Calderbank-Pedersen forms~\cite{Calderbank:2001uz}, which are controlled by a single function on the hyperbolic plane. The relation between the charged twistor spinor and the Killing spinor is given. We finally address the hyper-K\"ahler case and discuss the properties of gravitational instantons. 

This paper is organized as follows: the next section reviews and discusses the Euclidean PD family. We demonstrate that the Euclidean PD metric allows a conformally K\"ahler structure. In section~\ref{sec:KY}, we implement the coordinate transformations under which the self-dual PD metric collapses into the self-dual Carter family with opposite orientation. Section \ref{sec:SUSY} addresses the supersymmetry of self-dual PD metric. The hyper-K\"ahler case and properties of gravitational instantons are explored in section~\ref{sec:HK}. Finally, section \ref{sec:conclusion} concludes with several remarks. The appendix presents the bilinear classification of Euclidean supergravity solutions and shows the supersymmetry for the non-self-dual PD metric. The properties of type-D LeBrun solution is also discussed.


\section{Pleba\'nski-Demia\'nski solution}
\label{sec:PD}

In this paper we consider solutions in the Einstein-Maxwell-$\Lambda$ system described by the action 
\begin{align}
\label{action}
S=\frac{(-1)^{s+1}}{2\kappa_4^2} \int \D^4 x \sqrt{|g|} (
R -2\Lambda- F_{\mu\nu}F^{\mu\nu}) \,,  
\end{align}
where $ F$ is a Faraday 2-form and $\Lambda$ is a cosmological constant. 
We employ a convention of mostly plus sign for the metric and $s$ is the number of 
its negative eigenvalues, i.e., $s=1, 0$ respectively for Lorentzian and Euclidean signature. 
Bosonic equations of motion derived from the action (\ref{action}) read
\begin{align}
\label{EOM}
R_{\mu\nu } =\Lambda  g_{\mu\nu }+ 2 \left(F_{\mu\rho }F_\nu {}^\rho
 -\frac{1}{4}g_{\mu\nu }F_{\rho\sigma }F^{\rho\sigma }\right) \,, \qquad 
\D \star F=0 \,, \qquad \D F=0\,. 
\end{align}
The last equation can be locally solved  in terms of a one-form $A$ as $F=\D A$.

The PD metric \cite{PD} in the Lorentzian  Einstein-Maxwell-$\Lambda$ system 
describes the most general Petrov-D solution with an aligned non-null Maxwell field~\cite{DKM}, which reduces to the Kinnersley metric in the vacuum case~\cite{Kinnersley:1969zza}. The PD solution represents a uniformly accelerating
charged mass with rotation.  Physical interpretation and the limiting procedures to the subclass of the PD metric were explored in detail in \cite{Griffiths:2005qp}. The supersymmetry of the Lorentzian PD metric was discussed in \cite{KN}. 

In what follows, we restrict ourselves to the case with Euclidean signature. 
The Euclidean PD metric reads 
\begin{align}
\label{PD}
\D s^2 =&\frac 1{(1+b pq)^2} \left\{
\frac{Q(q)}{q^2-p^2} (\D \tau-p^2 \D \sigma)^2 +(q^2-p^2)\left(\frac{\D q^2}{Q(q)}-\frac{\D p^2}{P(p)}\right)+\frac{P(p)}{p^2-q^2} (\D \tau-q^2 \D \sigma)^2 
\right\}\,,
\end{align}
with 
\begin{align}
 \label{PD_A}
A=& \frac{Q_m p (\D \tau-q^2\D \sigma)
+Q_eq(\D \tau-p^2 \D \sigma)}{q^2-p^2}\,. 
\end{align}
The structure functions $P(p)$ and $Q(q)$ are quartic polynomials of their arguments and 
are given by 
\begin{align}
\label{PQ_fun}
P(p) =& \left[
({a^2-n^2}
-Q_e^2+Q_m^2)b^2-\Lambda/3 \right]p^4
-2b m p^3-p^2-2n p+{a^2-n^2} 
\,, 
\nonumber \\
Q(q) =& \left[({a^2-n^2})
b^2-\Lambda/3 \right]q^4+2 b n q^3-q^2+2mq+{a^2-n^2}
-Q_e^2+Q_m^2\,.  
\end{align}
For definiteness of the argument, we restrict to the coordinate range 
\begin{align}
\label{pqrange}
p^2-q^2>0\,,\qquad P(p)>0\,, \qquad  Q(q)<0\,, 
\end{align}
with the orientation $\epsilon_{\tau qp\sigma}>0$. 
The solution is parameterized by 7 constants ($m ,n, a, b, Q_m, Q_e, \Lambda$), which describe the mass, the NUT charge, the angular momentum, the acceleration, the magnetic charge, the electric charge and the cosmological constant.  The metric admits two commuting Killing vectors 
\begin{align}
\label{Killing}
\xi_{(1)}=\frac{\partial}{\partial \tau } \,, \qquad 
\xi_{(2)}=\frac{\partial}{\partial \sigma } \,,  
\end{align}
which correspond to the stationarity and axisymmetry in Lorentzian signature. 

The PD family incorporates a large variety of physically interesting spacetimes. Among them, the 
$b=0$ limit describes the Carter family~\cite{Carter:1968ks},\footnote{
In this paper, we do not discuss another interesting subclass called the C-metric~\cite{Kinnersley:1970zw}, 
since it does not admit a nontrivial self-dual limit. } for which 
the overall conformal factor $(1+bpq)^{-2}$ drops off. The Carter class of metrics admits a separability for geodesic, Klein-Gordon and Dirac equations~\cite{Carter:1968ks,Hughston:1972qf,Carter:1977pq} (see also \cite{Anabalon:2016hxg}). The responsibility for the integrability is borne by the nondegenerate rank-two KY  tensor, which gives rise to an additional constants of motion. The Euclidean Carter solution has been utilized in many contexts: 
ref. \cite{Behrndt:2002xm} constructed a new manifold with $G_2$ holonomy which has a topology $\mathbb R^3$ over the self-dual Carter metric, and  a new Sasaki-Einstein manifold was constructed from the Euclidean Carter base space in \cite{Martelli:2005wy}. 

\subsection{Conformally K\"ahler structure}

The Euclidean PD family is endowed with a variety of mathematically interesting structures. 
Taking the orthonormal frame by
\begin{align}
e^1& =\sqrt{\frac{Q(q)}{q^2-p^2}}
\frac{(\D \tau -p^2 \D \sigma )}{1+bpq}
\,, \qquad e^2 = \sqrt{\frac{q^2- p^2 }{Q(q)}} \frac{\D q}{1+bpq}
\,, \nonumber \\
e^3&= \sqrt{\frac{p^2-q^2}{P(p)}}\frac{\D p}{1+bpq} \,, \qquad 
e^4= \sqrt{\frac{P(p)}{p^2-q^2}}\frac{(\D \tau -q^2 \D
 \sigma )}{1+bp q} \,, 
 \label{basis}
\end{align}
we can define (anti-)self-dual two-forms
\begin{align}
\label{CS_PD}
J_\pm = e^1 \we e^2\pm e^3 \we e^4 \,, \qquad \star J_\pm =\pm J_\pm \,. 
\end{align}
Raising one index by the metric as $(J_\pm)_\mu{}^\nu =g^{\nu\rho}(J_\pm)_{\mu\rho}$, 
these tensors satisfy
\begin{align}
\label{}
(J_\pm)_\mu{}^\rho (J_\pm)_\rho{}^\nu=-\delta_\mu{}^\nu\,, \qquad 
(J_+)_\mu{}^\rho(J_-)_\rho{}^\nu=(J_-)_\mu{}^\rho(J_+)_\rho{}^\nu\,.
\end{align} 
Namely, these are commutative almost complex structures. Moreover, as shown in ref.~\cite{KN},  Nijenhuis tensors associated with these complex structures vanish\footnote{
When the manifold is compact, these complex structures are not defined globally, as exemplified by $S^4$ (see e.g.~\cite{Goldblatt:1994rx}). 
These mutually-commuting integrable almost complex structures are combined to give an 
integrable almost product structure 
$\Pi_\mu{}^\nu=(J_+)_\mu{}^\rho(J_-)_\rho{}^\nu$ 
satisfying $\Pi_{\mu\nu}=\Pi_{(\mu\nu)}$ \cite{Gauntlett:2003cy}. 
The integrability is measured by vanishing the Nijenhuis tensor 
$N[\Pi]_{\mu\nu}{}^\rho=\Pi_\mu{}^\sigma\partial_{(\sigma}\Pi_{\nu)}{}^\rho-\Pi_\nu{}^\sigma\partial_{(\sigma}\Pi_{\mu)}{}^\rho$. The vanishing of the tensor $N[\Pi]_{\mu\nu}{}^\rho$ is responsible for the metric to take the block-diagonal form.  
}
\begin{align}
\label{}
N[J_\pm]_{\mu\nu}{}^\rho\equiv J_{\pm\mu}{}^\sigma\partial_{[\sigma}J_{\pm\nu]}{}^\rho-J_{\pm\nu}{}^\sigma\partial_{[\sigma}J_{\pm\mu]}{}^\rho =0\,.
\end{align}
This is closely related to the fact that the PD metric is 
algebraically special~\cite{Goldblatt:1994rx}. 

In the case of a nonvanishing cosmological constant, the metric does not admit any covariantly constant 2-forms~\cite{Houri:2014hma}. It follows that the metric is not K\"ahler. It is then instructive to consider how it deviates from the K\"ahler metric. The obstruction for $J_\pm$ to be closed is ascribed to the presence of intrinsic torsion (see e.g, \cite{Chiossi:2002tw,Gauntlett:2003cy}). For the ${\rm SU}(2)\subset {\rm Spin}(4)$ structure, there appear at most three kinds of modules for intrinsic torsion. 
The PD family admits only a single kind of intrinsic torsion expressed in terms of the Lee form 
$\theta_\mu=\frac 12 J^{\nu\rho}(\D J)_{\mu\nu\rho}$ as 
\begin{align}
\label{Lee}
\D J_\pm = \theta _\pm \we J_\pm \,, \qquad 
\theta_\pm =\D \left[2 \log\left(\frac{q\mp p}{1+b p q}\right)\right] \,. 
\end{align}
Since these Lee forms are locally exact, the conformal transformations 
$\hat g_{\mu \nu}^\pm=\Omega_\pm ^2 g_{\mu\nu}$ 
with $\Omega_\pm =(1+bpq)/(q\mp p)$ lead to K\"ahler metrics with K\"ahler forms
$\hat J_\pm\equiv \Omega^2_\pm J_\pm$.  
Namely, the PD metric is conformally K\"ahler and the two K\"ahler metrics read
\begin{align}
\label{Kahler_conformalCarter}
\D s^2_\pm=\frac{1}{(q\mp p)^2} \left\{
\frac{Q(q)}{q^2-p^2} (\D \tau-p^2 \D \sigma)^2 +(q^2-p^2)\left(\frac{\D q^2}{Q(q)}-\frac{\D p^2}{P(p)}\right)+\frac{P(p)}{p^2-q^2} (\D \tau-q^2 \D \sigma)^2 
\right\}\,,
\end{align}
with each K\"ahler form locally given by
\begin{align}
\label{hatJpm}
\hat J_\pm=\D \left(\frac{\D \tau \mp pq \D \sigma}{q\mp p}\right)\,, 
\qquad 
\nabla _\mu^\pm \hat J_{\pm \nu\rho}=0 \,. 
\end{align}
$\nabla^\pm$ represent the Levi-Civita connection for $\hat g^\pm$. 
Both metrics (\ref{Kahler_conformalCarter}) remain K\"ahler independent of the precise form of structure functions $P(p)$ and $Q(q)$. Moreover, they are toric, namely they admit two commuting isometries~\cite{Abreu}. These toric K\"ahler geometries are conformally related to each other with opposite orientation, i.e., they are `ambi-K\"ahler' in the terminology of ref.~\cite{Apostolov:2013oza}. 
Recently, Gauduchon and Moroianu showed \cite{Gauduchon:2015} that in some situation a KY tensor gives rise to the ambi-K\"ahler structure. 

Let us make a deeper analysis of these K\"ahler spaces for later convenience. 
We refer to ref.~\cite{Apostolov:2013oza} for more concrete discussions. 
For more discussions on a Riemannian metric conformal to K\"ahler,
see \cite{Dunajski:2010}.
In what follows, we shall focus exclusively on 
$\hat g^-$.  Since the metric (\ref{Kahler_conformalCarter}) is toric, it falls into the canonical form of toric K\"ahler manifolds~\cite{Abreu}. Vectors (\ref{Killing}) still solve Killing's equation for $\hat g^-$ and are expressed as $\xi_{(I)}^\mu=-\hat J^{\mu\nu}_-\partial _\nu D_{(I)}$ ($I=1,2$), 
where $D_{(I)}$ are Killing prepotentials  given by
\begin{align}
\label{}
D_{(1)}=\frac{1}{q+p} \,, \qquad D_{(2)}=\frac{pq}{p+q}\,. 
\end{align}
These prepotentials are commutative under the Poisson bracket induced by $\hat J^{\mu\nu}_-$. 
If we define the symplectic metric $\mathbb G_{IJ}$ by
\begin{align}
\label{}
\mathbb G_{IJ}\D D_{(I)} \D D_{(J)}=\frac{p+q}{p-q}
\left[
\frac 1{P(p)}
(\D D_{(2)}+p^2 \D D_{(1)})^2-\frac{1}{Q(q)}(\D D_{(2)}+q^2 \D D_{(1)})^2 \right]\,,
\end{align}
the metric can be cast into the desired form 
\begin{align}
\label{}
\D s^2_-=\mathbb G_{IJ}\D D_{(I)}\D D_{(J)} +
\mathbb G^{IJ}\D \phi_I \D \phi_J\,, 
\end{align}
where $\phi_I=(\tau, \sigma)$ and $\mathbb G^{IJ}$ is the inverse of $\mathbb G_{IJ}$. 
The Ricci form 
$\hat{\mathfrak R}_{\mu\nu}=\frac 12 \hat R^{\rho\sigma}{}_{\mu\nu}\hat J_{-\rho\sigma}$ is  given by 
\begin{align}
\label{RF}
\hat{\mathfrak R}
=& \D \hat{W}\,, \qquad 
\hat{W}=
\frac{-(p+q)Q'+4Q}{2(p+q)^2(p-q)}(\D \tau -p^2 \D \sigma) 
+ \frac{(p+q)P'-4P}{2(p+q)^2(p-q)}(\D \tau -q^2 \D \sigma) 
\,, 
\end{align}
where the prime denotes the partial differentiation with respect to the corresponding arguments. 
The K\"ahler metric admits a constant spinor $\varepsilon$ satisfying 
\begin{align}
\label{}
(\nabla^-_\mu +\hat{W}_\mu)\varepsilon =0\,, \qquad 
\frac 14 \hat J^-_{\mu\nu} \hat \gamma^{\mu\nu} \varepsilon =-i \varepsilon\,. 
\end{align}

We have seen that both the PD metric and the Carter metric ($b=0$) are conformal to the 
K\"ahler metric (\ref{Kahler_conformalCarter}). This fact turns out helpful 
when we try to obtain the Killing spinors of self-dual PD metric as will be discussed in 
section \ref{sec:SUSY}.

\subsection{(Anti-)self-dual case}

Let us consider the case where the Weyl curvature is (anti-)self-dual 
\begin{align}
\label{SD_W}
C_{\mu\nu\rho\sigma}=\pm \frac 12 \epsilon_{\mu\nu\tau\lambda}C^{\tau\lambda}{}_{\rho\sigma} \,.
\end{align} 
Equivalently, the Weyl tensor is referred to as half conformally flat.  
For the PD metric, this condition is satisfied provided the parameters obey the following restrictions~\cite{KN} 
\begin{align}
\label{SD_PD}
m=\pm n \,, \qquad 
Q_e=\pm Q_m \,.
\end{align}
The second condition is nothing but the (anti-)self-duality of the 
Maxwell field, i.e, 
\begin{align}
F_{\mu\nu}=\pm \frac 12\epsilon _{\mu\nu\rho\sigma}F^{\rho\sigma} \,. 
\label{SD_F}
\end{align}
In this case, the stress-tensor of the Maxwell field vanishes, yielding an Einstein space. 
In fact, the (anti-)self-dual Maxwell field is proportional to the K\"ahler form
\begin{align}
\label{}
F=\frac{Q_e(1+bpq)^2}{(p\mp q)^2}J_\pm \,. 
\end{align}

The (anti-)self-dual PD metric is therefore characterized by four parameters ($m, a, b ,\Lambda$). 
We shall focus on the self-dual PD metric in what follows. 
The self-duality of the Weyl tensor implies that  the holonomy of the manifold is contained in ${\rm Sp}(1)\times {\rm Sp}(1)$, viz, it is a quaternionic-K\"ahler manifold~\cite{FVP}. 
Taking the anti-self-dual two forms
\begin{align}
\label{Ji_ASD}
J^{(1)}=J_- \,, \qquad J^{(2)}=e^1\we e^3-e^4\we e^2 \,, \qquad 
J^{(3)}= e^1\we e^4-e^2\we e^3\,,
\end{align}
which are subjected to the imaginary unit quaternions algebra
$J^{(i)}\cdot J^{(j)}=-\delta^{ij}+\epsilon_{ijk}J^{(k)}$, 
one can check that the characterization of a quaternionic manifold 
\begin{align}
\label{}
\mas F^{(i)}=\frac 13 \Lambda J^{(i)}\,, \qquad 
\mathfrak R^{(i)} =\mas F^{(i)} \,, 
\end{align}
is fulfilled, where 
$\mathfrak R^{(i)}{}_{\mu\nu}=\frac 12R^{\rho\sigma}{}_{\mu\nu}J^{(i)}{}_{\rho\sigma}$
are the Ricci forms and 
$\mas F^{(i)}$ are the ${\rm SU}(2)$ curvatures defined by
\begin{align}
\label{}
 \mas F^{(i)}=\D \mas A^{(i)} +\epsilon_{ijk}\mas A^{(j)} \we \mas A^{(k)} \,, \qquad 
 \mas A^{(i)}{}_\mu\equiv \frac 12 \Omega_{\mu ab}J^{(i)ab} \,. 
\end{align}
Here, 
$\Omega_{\mu ab}=e_{a\nu}\nabla_\mu e_b{}^\nu$ is the spin connection. 
$J^{(i)}$ are covariantly constant with respect to the ${\rm SU}(2)$ connections
\begin{align}
\label{}
\mas D_\mu J^{(i)}{}_{\nu\rho}\equiv \nabla_\mu J^{(i)}{}_{\nu\rho}+\epsilon_{ijk}
\mas A^{(j)}{}_\mu J^{(k)}{}_{\nu\rho} =0 \,. 
\end{align}

It has been recently shown in ref.~\cite{Houri:2014hma} that 
when the Weyl tensor is self-dual, 
the PD metric admits a nondegenerate KY tensor 
\begin{align}
\label{KY_PD}
f = -f_1 e^1 \we e^2+ f_2 e^2 \we e^4+f_2 e^1\we e^3+f_3 e^3 \we e^4 \,, 
\end{align}
where 
\begin{align}
\label{f123}
f_1 =\frac{b f_0 (p+q)^{-1} +\Lambda p}{1+b pq} \,, \qquad 
f_2 = \frac{3 b (p-q)\sqrt{-P(p ) Q(q)} }{(1+b p q)(p^2-q^2)} \,, \qquad 
f_3 =\frac{b f_0 (p+q)^{-1} +\Lambda q}{1+b pq} \,, 
\end{align}
with
\begin{align}
\label{}
f_0=-3 (a^2-m^2) [(1+b p q)^2+b(p^2+q^2)]-3m(1-bpq)(q-p)+\Lambda p^2 q^2-3 p q\,. 
\end{align}
The two-form $f_{\mu\nu}$ satisfies the KY equation~\cite{Yano}
\begin{align}
\label{}
\nabla_{(\mu } f_{\nu)\rho} =0 \,. 
\end{align}
When the acceleration parameter is turned off ($b=0$), the KY tensor (\ref{KY_PD}) reduces to the known one found by Floyd-Penrose~\cite{Floyd,Penrose:1973um}. 
The existence of the above KY tensor seems surprising, since it would give a constant of motion due to which the geodesic motion is integrable.  In the Lorentzian case, the conformal factor of the general PD metric is a main obstacle to separate variables and no rank-two KY tensor exists in that case~\cite{Houri:2014hma}. An important fact here is that the KY tensor (\ref{KY_PD}) is nondegenerate. In this case, the theorem given in ref.~\cite{Houri:2007xz} determines the local metric to the Carter family ($b=0$), whereas the selfdual PD metric appears to fall out of this class of metrics. Hence, the consistency to the theorem in~\cite{Houri:2007xz} and the origin of this additional hidden symmetry remain obscure.  To fill this gap is one of the main results of the present paper. 
 
When the cosmological constant is set to zero, the self-duality condition for the Weyl tensor (\ref{SD_PD})
implies the self-duality of the Riemann tensor. In this case, the metric is hyper-K\"ahler and 
the KY tensor (\ref{KY_PD}) is covariantly constant, thence
is proportional to the hyper-K\"ahler form.  The details are shown in section \ref{sec:HK}.

It would be worth discussing whether the (anti-)self-dual PD metric is in the conformal class of a Riemannian metric with the vanishing Ricci tensor. The motivation comes from the conformal gravity whose action is given by the squared norm of the Weyl tensor. The equations of motion are the vanishing of the Bach tensor, which are invariant under conformal transformations. It is known that a metric conformal to a Ricci-flat metric is Bach-flat, i.e., a solution of the conformal gravity theory. Less known is that a Riemannian metric with (anti-)self-dual Weyl curvature is Bach-flat. In \cite{Dunajski:2014}, necessary and sufficient conditions for a Riemannian metric with anti-self-dual Weyl tensor to be locally conformal to a Ricci-flat metric were provided. They are, in terms of the Weyl tensor $C_{\mu\nu\rho\sigma}$ and the Schouten tensor $P_{\mu\nu}=- (1/2)R_{\mu\nu}+ (1/12)g_{\mu\nu} R $, given by
\begin{align}
 4 \nabla^\sigma C_{\mu\nu\rho\sigma} \nabla_\kappa C^{\mu\nu\rho\kappa}
 - |V|^2|C|^2 = 0 \,, \qquad
 P_{\mu\nu} + \nabla_\mu V_\nu + V_\mu V_\nu - \frac{1}{2} |V|^2 g_{\mu\nu} = 0 \,,
\end{align}
where $V_\mu = (4/|C|^2) C^{\nu\rho\sigma}{}_\mu \nabla^\kappa C_{\nu\rho\sigma\kappa}$,
$|C|^2=C_{\mu\nu\rho\sigma}C^{\mu\nu\rho\sigma}$ and $|V|^2=V_\mu V^\mu$.
Since the anti-self-dual PD metric does not satisfy both conditions,
its conformal class does not contain a Ricci-flat metric.
For other examples of anti-self-dual metrics without Ricci-flat metrics in their conformal class, see \cite{Dunajski:2014}.

\section{Killing-Yano tensor of self-dual Pleba\'nski-Demia\'nski solution}
\label{sec:KY}

In this section, we explore how the unexpected KY tensor (\ref{KY_PD}) arises in the self-dual PD family. 
The additional symmetry represented by the KY tensor allows us to show that the self-dual PD family is isometric to the self-dual Carter family. We explore the desired coordinate transformation utilizing the KY tensor (\ref{KY_PD}).

\subsection{Anti-self-dual conformal Killing-Yano tensor}

Let us consider the general Einstein space with a self-dual Weyl curvature that possesses a 
Killing vector $\xi$. In terms of the Killing one-form,  we can define an anti-self dual two-form
\begin{align}
\label{k_CKY}
k[\xi]=\frac {1}2 (\D \xi -\star \D \xi) \,, \qquad \star k =- k \,. 
\end{align}
Here and in what follows, we employ the notation $k[\xi]$ to illustrate which Killing vectors 
are considered, and we omit it when no confusions arise. From the integrability condition for the Killing vector 
$\nabla_\mu\nabla_\nu \xi_\rho =R_{\rho\nu\mu\sigma}\xi^\sigma$, 
it is easy to check that $k$ is an anti-self dual 
CKY tensor, satisfying the CKY equation
\begin{align}
\label{CKYeq}
\nabla_{(\mu} k_{\nu) \rho } =\frac 13\big(\nabla^\sigma k_{\sigma\rho}g_{\mu\nu} -
 \nabla^\sigma k_{\sigma(\mu }g_{\nu)\rho}  \big)\,, \qquad 
\nabla^\nu k_{\nu\mu}[\xi] =-\Lambda \xi_\mu \,. 
\end{align}
Note that in Einstein spaces the divergence of the CKY tensor 
always gives rise to a Killing vector.
For the PD metric, we have two independent ${\rm U}(1)$ Killing vectors
$\xi_{(1)}=\partial/\partial\tau$ and $\xi_{(2)}=\partial/\partial\sigma$, so that we can 
construct two kinds of anti-self dual CKY tensors as above.

Since the general PD metric belongs to Petrov-D space, 
it admits another CKY tensor
\begin{align}
\label{CKY_Y}
Y= \frac 1{1+bpq} ( p e^1 \we e^2 -q e^3 \we e^4) \,, \qquad 
\nabla_\nu Y{}^{\nu\mu} = -3b 
(\partial/\partial\sigma)^\mu \,.
\end{align}
For $b=0$, this reduces to the the KY tensor found by Floyd and Penrose~\cite{Floyd,Penrose:1973um}. 
Here, let us decompose it into the self-dual and anti-self-dual parts 
\begin{align}
\label{}
Y_\pm =\frac 12 (Y \pm \star Y) \,, \qquad \star Y_\pm =\pm Y_\pm \,, \qquad 
\nabla_\nu Y_\pm {}^{\nu\mu}= \frac 32
\left(\pm\frac{\partial}{\partial \tau}-b\frac{\partial}{\partial \sigma}\right)^\mu
\,.
\end{align}
Since the Hodge dual of a CKY is necessarily a CKY tensor, both of $Y_\pm$ satisfy the CKY equation. 
These two tensors are proportional to the almost K\"ahler forms given in (\ref{CS_PD}):
\begin{align}
\label{kpm_Jpm}
Y_\pm=  \frac{p\mp q}{2(1+bpq)}J_\pm \,. 
\end{align}
The condition that $Y_\pm$ satisfy the CKY equation is due to the fact that K\"ahler forms $\hat J_\pm$ (\ref{hatJpm}) are  covariantly conserved with respect to the Levi-Civita connection of $\hat g^\pm$. 
Note that $Y_-$ is not entirely new, since it is simply given by
\begin{align}
\label{}
Y_-=\frac 3{2\Lambda}k[
\partial/\partial\tau +b\partial/\partial\sigma
] \,.
\end{align}
Hence the information for the Petrov-type is encoded in $Y_+$.

It turns out that  there appear three independent CKY tensors
($k[\partial/\partial\tau]$, 
$k[\partial/\partial\sigma]$, $Y_+$) in the self-dual PD metric. 
None of these tensors are closed nor coclosed. Nevertheless, 
a particular linear combination of these CKY tensors 
gives a {\it closed} CKY tensor
\begin{align}
\label{}
h \equiv 
\frac 32 
k[\partial/\partial\tau-b\partial/\partial\sigma ]-\Lambda 
Y_+\,, \qquad \D h=0 \,, \qquad 
\nabla_\nu h^{\nu\mu} =3\Lambda (-\partial/\partial\tau+b\partial/\partial\sigma)^\mu \,. 
\end{align}
The Hodge-dual of the closed CKY tensor 
gives a KY tensor~\cite{Yasui:2011pr}
\begin{align}
\label{}
f= \star h =-\frac 32 k[\partial/\partial\tau-b\partial/\partial\sigma] -\Lambda Y_+\,. 
\end{align}
One can verify that this recovers the one given in~(\ref{KY_PD}).

We have illustrated how the KY tensor (\ref{KY_PD}) is derived in the present settings. 
From the perspective of general PD family, the anti-self-dual CKY tensors~(\ref{k_CKY}) are additional symmetries, whereas the CKY tensor (\ref{CKY_Y}) resident in type D metric is an unexpected symmetry from the viewpoint of manifolds with a self-dual Weyl curvature.  As will be discussed in section~\ref{sec:twistor}, the CKY tensor~(\ref{k_CKY}) bears a consistent relevance to the twistor spinor.

\subsection{Transformation to the Carter family}
\label{sec:Carter}

According to the analysis in ref.~\cite{Houri:2007xz}, 
the existence of a closed CKY 2-form, which is dual to the KY two-form, locally determines the 
metric to the Carter family. Hence, there must exist a coordinate transformation which brings the self-dual PD metric into the contracted Carter metric corresponding to $b=0$. At first sight, this might be a formidable task since the metric nontrivially depends on the variables ($p, q$). In order to find the desired coordinate transformation, the nondegenerate KY tensor (\ref{KY_PD}) plays a crucial role. To this end, let us consider the eigenvalue problem of the associated Killing tensor~\cite{Krtous:2006qy}: 
\begin{align}
\label{}
K^a{}_b v^b =\lambda ^2 v^a \,, \qquad K^a{}_b \equiv -f^a{}_cf^c{}_b \,. 
\end{align}
In order to diagonalize $K^a{}_b$, let us introduce another orthonormal frame $e^{(i)}$ 
with the transformation  matrix $E^a{}_{(i)}$ by $e^a=E^a{}_{(i)}e^{(i)}$. Denoting the inverse of $E^a{}_{(i)}$ as $E^{(i)}{}_a$, we get 
$K^a{}_b=E^a{}_{(i)}K^{(i)}{}_{(j)}E^{(j)}{}_b$, where $K^{(i)}{}_{(j)}$ enjoys only the 
diagonal entities. The four eigenvalues are pairwise equal and are given by 
\begin{align}
\label{}
\lambda_\pm =\frac 12 \left(f_1-f_3 \pm\sqrt{4f_2^2+(f_1+f_3)^2}\right) \,,
\end{align}
where $f_i$ is given by (\ref{f123}). The transformation matrix $E^a{}_{(i)}$ reads
\begin{align}
\label{}
E^a{}_{(i)}=\left(
\begin{array}{cccc}
 A_+/B_+  & 0  &   0& A_-/B_-    \\
0   & -A_+/B_+ & -A_-/B_-  &0    \\
 0  & 1/B_+ & 1/B_-  &  0  \\
  1/B_+ & 0 &0   & 1/B_-   
\end{array}
\right)\,, 
\end{align}
where 
\begin{align}
\label{}
A_\pm =\frac{f_1+f_3\pm\sqrt{4f_2^2+(f_1+f_3)^2}}{2 f_2}\,, \qquad 
B_\pm =\sqrt{1+A_\pm ^2} \,. 
\end{align}
It is worth emphasizing  the property 
\begin{align}
\label{frame_rel}
{\rm det}(E^a{}_{(i)})=-1 \,. 
\end{align}
This means that the orientation of the frame $e^{(i)}$ is reversed 
from that of $e^{a}$. For instance, the self-dual almost K\"ahler form (\ref{CS_PD}) transforms 
into $J_+=-e^{(1)}\we e^{(2)} +e^{(3)}\we e^{(4)} $ with 
$e^{(1)}\we e^{(2)}\we e^{(3)}\we e^{(4)}$ being negatively oriented.  This property is of relevance to the ambi-K\"ahler structure and is essential for obtaining the supersymmetric canonical forms.

In the new frame $e^{(i)}$ it turns out that the KY tensor (\ref{KY_PD}) takes the following form
\begin{align}
\label{}
f= \lambda_+ e^{(1)} \we e^{(2)} -\lambda_- e^{(3)}\we e^{(4)} \,.
\end{align}
This is exactly the same form as the KY tensor in the Carter family~\cite{Frolov:2008jr,Yasui:2011pr}. 
This fact prompts us to employ the eigenvalues as new coordinates $(p,q)\to (u,v)$:
\begin{align}
\label{pq_uv}
u=-\lambda_- (p,q)\,, \qquad v=\lambda _+(p,q) \,.
\end{align}
These equations can be inversely solved to give 
\begin{subequations}
\label{qpsoluv}
\begin{align}
q&=\frac{(-u+v)N_1\pm \sqrt{(u-v)^2N_1^2-4 N_2(bN_2-\Lambda N_1)}}{2(bN_2-\Lambda N_1)}\,, \\
p&=\frac{(u-v)N_1\pm \sqrt{(u-v)^2N_1^2-4 N_2(bN_2-\Lambda N_1)}}{2(bN_2-\Lambda N_1)}\,, 
\end{align}
\end{subequations}
where
\begin{align}
\label{}
N_1&=6(a^2-m^2)b^2+3b-\Lambda \,, \notag \\
N_2&=18(a^2-m^2)^2b^3+
3(a^2-m^2)b(3b-\Lambda)+(3bm-u)(3bm+v) \,. 
\end{align}
The sign choice in (\ref{qpsoluv}) is fixed depending on $\Lambda N_1(1+bpq)(p+q)\gtrless 0$. 
We shall consider the upper sign for definiteness. 
Together with a ${\rm GL}(2,\mathbb R)$ transformation of Killing coordinates
\begin{align}
\label{tausigma_tr}
\left(\begin{array}{c}
\tau\\ \sigma 
\end{array}\right)
=
\left(\begin{array}{cc}
\Lambda & -3 b [(a^2-m^2)N_1+3bm^2]\Lambda \\ 
-b\Lambda & [\{3 (a^2-m^2)b^2-\Lambda\}N_1+9b^3m^2]\Lambda 
\end{array}\right)
\left(\begin{array}{c}
\psi\\ \chi
\end{array}\right)\,, 
\end{align}
somewhat lengthy but straightforward calculations show that the self-dual PD metric is converted into the following form
\begin{align}
\label{Carter}
\D s^2= \frac{\ma Q(u)}{u^2-v^2}(\D \psi-v^2 \D \chi)^2 
+\frac{u^2-v^2}{\ma Q(u)}\D u^2   +\frac{v^2-u^2}{\ma P(v)}\D v^2   
+ \frac{\ma P(v)}{v^2-u^2}(\D \psi-u^2 \D \chi)^2 \,, 
\end{align}
where 
\begin{align}
\label{Carter_PQ}
\ma Q(u)=-\frac{\Lambda}3u^4+\alpha_2 u^2+2\alpha_1 u+\alpha_0 \,, \qquad 
\ma P(v)=-\frac{\Lambda}3 v^4+\alpha_2 v^2 -2 \alpha_1 v+\alpha_0 \,,
\end{align}
with 
\begin{subequations}
\begin{align}
\label{}
   \alpha_0 =&-27 b^4[a^2+2(a^2-m^2)^2 b]^2\Lambda+
   9b^2[1+4(a^2-m^2)b][a^2+2b(a^2-m^2)^2]\Lambda^2
  \nonumber \\ &  -3 b[2(a^2-m^2)+5(a^2-m^2)^2b+2m^2]\Lambda^3+(a^2-m^2)\Lambda^4  \,, \\
   \alpha_1 =&m\Lambda^3 \,,\\
  \alpha_2 =  & 6b^2[2b(a^2-m^2)^2+a^2]\Lambda -[1+4(a^2-m^2)b]\Lambda^2 \,.   
\end{align}
\end{subequations}
In the new coordinates, the gauge potential~(\ref{PD_A}) is given by
\begin{align}
\label{A_uv}
A=\frac{Q_e \Lambda^2}{u-v}(\D \psi-uv\D \chi) \,,  
\end{align}
where the pure gradient piece was gauged away.  In the parameter region 
$\Lambda<0$, $N_1>0$ together with (\ref{pqrange}), the new frame $e^{(i)}$ is given by 
\begin{align}
\label{Carter_frame}
e^{(1)}&=-\sqrt{\frac{\ma Q(u)}{u^2-v^2}}(\D \psi-v^2 \D \chi) \,, \qquad 
e^{(2)}=\sqrt{\frac{u^2-v^2 }{\ma Q(u)}}\D u\,, \notag \\
e^{(3)}&=\sqrt{\frac{u^2-v^2 }{-\ma P(v)}}\D v\,, \qquad 
e^{(4)}=\sqrt{\frac{-\ma P(v)}{u^2-v^2}}(\D \psi-u^2 \D \chi) \,. 
\end{align}
The derived solution (\ref{Carter}) incorporates four parameters ($\alpha_0, \alpha_1, \alpha_2, \Lambda$), which is the same number as the self-dual original PD solution. 
However, there exists an additional scaling symmetry 
$(u,v,\psi, \chi )\to  (\lambda u,\lambda v, \lambda^{-3}\psi, \lambda^{-5}\chi )$
with $(\Lambda, b, m+i a) \to (\lambda^2\Lambda, \lambda^2 b, \lambda^{-1}(m+ia))$, 
which allows us to set $\alpha_2=-1$ without losing any generality. The resultant metric  (\ref{Carter}) is the Carter solution obtained by the $b\to 0$ limit of the self-dual PD solution. 
Since the current coordinate transformation reverses the orientation (\ref{frame_rel}), 
$e^{(1)}\we e^{(2)}\we e^{(3)}\we e^{(4)}$ must be {\it negatively} oriented. 
With this remark, the metric (\ref{Carter}) is the Carter family with a self-dual Weyl curvature. Note that this statement is not valid for the Ricci-flat case, since the KY tensor (\ref{KY_PD}) is degenerate in that case and its eigenvalues cannot be used as new coordinates. 

Let us conclude this section by commenting why the acceleration parameter $b$ can be eliminated. 
In the general case with a self-dual Weyl curvature, the only independent components of the Weyl tensor are encoded into the electric part, corresponding to the five independent element of $C_{4i4j}$ ($i,j=1,2,3$). In the case of the self-dual PD family, the curvature is algebraically special so that only a single independent component remains nonvanishing
\begin{align}
\label{Psi2PD}
\Psi_{0011}^+\equiv -C_{\mu\nu\rho\sigma}l^\mu m^\nu \bar m^\rho \bar l^\sigma =
-\frac{2m(1+bpq)^3}{(p-q)^3}, 
\end{align}
where $l_\mu=(e^1+i e^2)_\mu/\sqrt 2$, $m_\mu=(e^3+i e^4)_\mu/\sqrt 2$ with the basis given by (\ref{basis}). Hence, the particular combination of variables appearing in $\Psi_{0011}^+$ is physically meaningful. This is in contrast to the Lorentzian Petrov-D metrics, in which two real curvature components remain independent. This fact effectively eliminates the acceleration parameter $b$. Actually, if we insert (\ref{qpsoluv}),  we get $\Psi_{0011}^+=2m\Lambda^3/(u-v)^3$, which precisely recovers the one for the self-dual Carter family. A similar situation occurs in five-dimensional Kerr-NUT-de Sitter metrics, in which a specific combination of the mass and the nut charge enters in the curvature invariants. Therefore,  one of these parameters can be set to zero by a suitable coordinate transformation~\cite{Chen:2006ea}.



\section{Supersymmetry of Pleba\'nski-Demia\'nski solution} 
\label{sec:SUSY}

The localization technique in Euclidean conformal field theory~\cite{Martelli:2012sz,Martelli:2013aqa,Farquet:2014kma,Hama:2011ea,Hosomichi:2015jta} is a main driving force behind the study of supersymmetry in Euclidean signature.  
Since the phenomenon of localization allows an exact nonperturbative computations just by saddle point computations, the Euclidean supersymmetry has become a focus of attention.  Inspired by this issue, some authors have classified Euclidean supersymmetric backgrounds~\cite{Dunajski:2010zp,Gutowski:2010zs,Dunajski:2010uv}.  In this section, we demonstrate that the self-dual PD metric indeed admits two-independent Killing spinors. 

The supersymmetric solutions in the Euclidean Einstein-Maxwell-$\Lambda$ system (\ref{action}) is characterized by the existence of a Killing spinor $\epsilon$, which obeys a first-order differential equation~\cite{Dunajski:2010zp,Gutowski:2010zs,Dunajski:2010uv}
\begin{align}
\label{SCD}
\hat\nabla_\mu\epsilon \equiv \left(\nabla_\mu +\frac{i}{4}F_{\nu\rho }\gamma^{\nu\rho }\gamma_\mu
  - i \sqrt{\frac{-\Lambda}{3}}
A_\mu +\frac{1}{2}\sqrt{\frac{-\Lambda}{3}}\gamma_\mu \right)\epsilon=0 \,. 
\end{align}
In this section, we restrict ourselves mostly to the $\Lambda=-3\ell^{-2}<0$ case.\footnote{
In Lorentzian signature, the positive cosmological constant is not compatible with supersymmetry. 
In Euclidean signature, on the other hand, the $\Lambda>0$ case is not ruled out for physical reasons. The restriction to the $\Lambda<0$ case in this paper is due to technical reasons. See the discussion at section~\ref{sec:conclusion}.
} 
We relegate the classification of BPS solutions by bilinear method to appendix~\ref{app:SUSY_PD}. The canonical form of the BPS metric and the gauge field can be written as (\ref{BPSmetric_AdS}) and (\ref{Asol}), both of which are determined by solving the nonlinear system (\ref{BPSsystem}). Once these equations are solved, 
the twist one-form $\omega$ of the bilinear Killing field 
$U^\mu \equiv i\epsilon ^\dagger \gamma_5 \gamma^\mu \epsilon=(\partial/\partial t)^\mu$
is obtained by (\ref{domega}) and the Killing spinor is given by (\ref{sol_KS}), i.e., the solution preserves at least one quarter of supersymmetry. 

The main difficulty in finding BPS solutions in gauged supergravity lies in the nonlinearity of governing equations~(\ref{BPSsystem}). Without the classification scheme, one can sometimes integrate the Killing spinor equation directly.  In the non-static case, however, solving the Killing spinor equation with $\Lambda$ is a formidable task since the equations are dependent non-trivially on the radial and angular coordinates. 
It has been shown only recently in ref.~\cite{KN} that the Kerr-Newman-AdS metric indeed admits the Killing spinor, which illustrates this difficulty. 

The procedure employed in~\cite{KN} for obtaining the Killing spinor of the Lorentzian PD metic is two-fold. The first step is to obtain the integrability condition ${\rm det}[\hat \nabla_\mu, \hat \nabla_\nu]=0$, which provides the necessary conditions for the existence of a Killing spinor. 
Under this condition, one next seeks a Killing vector which is always timelike outside the horizon. 
This is an obvious candidate of a Killing vector  constructed from a Killing spinor, since 
it cannot be spacelike throughout the Lorentzian manifold. 
For the Lorentzian PD family, this constitutes the linear combination of two Killing fields (\ref{Killing}). 
This expectation is indeed true and it was confirmed in \cite{KN} that one can construct tensorial bilinears satisfying all the bilinear equations. It follows that the metric can be brought to the canonical form of BPS metric, and the solution of 
the Killing spinor is easily found.  As shown in appendix~\ref{app:SUSY_PD}, the Killing spinor for the non-self-dual PD metic can be obtained by a suitable Wick rotation of the metric in the Lorentzian signature. Unfortunately, the bilinears do not have a well-defined self-dual limit, in which case a separate analysis is required.  In this section, we explore the supersymmetry of self-dual PD metric, which has no analogue in the Lorentzian world.

Suppose that the Weyl tensor and the Maxwell field are self-dual, i.e., 
assume the conditions (\ref{SD_W}) and (\ref{SD_F}).  It has been widely recognized that some self-dual metrics give rise to the supersymmetric backgrounds. For either sign of the 
cosmological constant, the supercovariant derivative (\ref{SCD}) in the self-dual case thus simplifies to
\begin{align}
\label{KS_SD}
\hat \nabla_\mu \epsilon 
= \left(
\nabla_\mu -i \sqrt{-\frac{\Lambda }{3}}A_\mu+\frac 1{2}\sqrt{-\frac{\Lambda }{3}} \gamma_\mu 
-\frac {i} {2}F_{\mu\nu}(1-\gamma_5) \gamma^\nu 
\right)\epsilon \,.
\end{align} 
Assuming the bosonic equations of motion~(\ref{EOM}), 
the integrability condition for the Killing spinor $0=[\hat \nabla_\mu, \hat \nabla_\nu]\epsilon$ reduces to  
\begin{align}
\label{KS_SD_int}
[\hat \nabla_\mu, \hat \nabla_\nu]\epsilon =(1-\gamma_5 )\left[
\frac{1}{8}C_{\mu\nu\rho\sigma }\gamma^{\rho\sigma }+\frac{i}{2}
\nabla_\rho F_{\mu\nu }\gamma^\rho -i \sqrt{-\frac{\Lambda }{3}}
\left(F_{\mu\nu }+\frac{1}{2}\gamma_{[\mu }{}^\rho F_{\nu ]\rho }\right)
\right]\epsilon =0\,. 
\end{align}
It then follows that ${\rm det}[\hat \nabla_\mu, \hat \nabla_\nu]=0$
is always satisfied, due to the overall projection factor $1-\gamma_5$. 
This means that the self-dual metric in the Einstein-Maxwell-$\Lambda$ system always fulfills the 
first integrability condition for the existence of a Killing spinor. The same conclusion is true for the 
anti-self-dual case. Since the integrability condition ${\rm det}[\hat \nabla_\mu, \hat \nabla_\nu]=0$ is merely a necessary condition for the preservation of supersymmetry, 
there may be a possibility that these are not sufficient (see e.g \cite{Cacciatori:2004rt} for a concrete example). To demonstrate rigorously that these are indeed sufficient we are forced to construct Killing spinors.

\subsection{Transformations to the Przanowski-Tod metric}

As reviewed in appendix~\ref{app:SUSY_PD}, the self-dual supersymmetric metrics for $\Lambda<0$
are represented by the Przanowski-Tod form~\cite{Przanowski:1991ru,Tod:2006wj}  
\begin{align}
\label{PT}
\D s^2=\frac{\ell^2}{z^2} \left[
H^{-1} (\D t +\omega)^2 +H \{\D z^2+e^\varphi (\D x^2+\D y^2)\} 
\right] \,, 
\end{align}
where 
$\varphi$ obeys the ${\rm SU}(\infty)$ Toda equation
\begin{align}
\label{Toda}
\Delta \varphi +\partial_z ^2 (e^\varphi)=0 \,.
\end{align}
Here $\Delta=\partial_ x^2+\partial_y^2$ is the two-dimensional Laplacian. 
$ H$ and $\omega $ are given in terms of  $\varphi$ as 
\begin{align}
\label{PT_omega}
\D \omega =\partial_x  H \D y \we \D z-\partial_y H \D x \we \D z +\partial_z (e^\varphi H)\D x \we \D y\,, 
\qquad 
H=1 -\frac 12 z\partial_z \varphi \,. 
\end{align}
The metric in the square bracket of (\ref{PT}) is the LeBrun space~\cite{LeBrun}, which is the scalar-flat K\"ahler space with an anti-self-dual K\"ahler form. Once the continuous Toda equation (\ref{Toda}) is solved, the Killing spinor is given by (\ref{KSsol_SD}), which preserves at least one quarter of supersymmetry. 
Although the Toda equation is known to be integrable, it seems intractable in practice to find explicit solution of the Toda equation. Moreover, it is far from clear which solutions of the Toda equation (\ref{Toda}) lead to the self-dual PD metric. Thus, here we do not try to directly solve the Toda equation. Instead, 
our strategy is to rewrite the self-dual PD metric into the Przanowski-Tod metric (\ref{PT}), and check all the  bilinear equations given in appendix~\ref{app:SUSY_PD}. Moreover, we can expect that the self-dual PD metric is a half BPS solution, since the matrix $[\hat \nabla_\mu, \hat \nabla_\nu]$ is rank two, i.e., it has two zero eigenvalues. To find two independent Killing spinors, one has to write the PD metric into Przanowski-Tod form (\ref{PT}) in two different manners. 

In the Lorentzian case, we can find the likely candidate of the supersymmetric bilinear Killing vector field of PD family,  by taking the linear combination of the two commuting Killing fields and looking for the coefficients which make the Killing vector everywhere timelike~\cite{KN}.  Unfortunately, this procedure does not work in Euclidean signature, since the vector fields are always spacelike. In the present case, it is much more prospective to first find the coordinate $z$, which appears as a conformal factor relating the Przanowski-Tod metric (\ref{PT}) and the LeBrun metric (\ref{LeBrun}).   

Now the PD metric is an Einstein space with a constant scalar curvature $R=4\Lambda$
and the LeBrun metric (\ref{LeBrun}) has a vanishing scalar curvature,  it turns out that 
the conformal factor $z$ must obey the 
conformal scalar equation on the Przanowski-Tod space
\begin{align}
\label{CCeq}
\nabla^\mu\nabla_\mu z -\frac 16 R z =0 \,.  
\end{align}
We follow the construction by Tod~\cite{Tod:2006wj} to find the solution to this equation. Let us denote a linear combination of two ${\rm U}(1)$ Killing vectors of PD metric~(\ref{Killing}) as 
\begin{align}
\label{U_c1c2}
U= c_1 \xi_{(1)} +c_2 \xi_{(2)} \,, 
\end{align}
where $c_1$ and $c_2$ are constants. Using the Killing vector above, 
one can define an anti-self-dual CKY two-form $k[U]$ as (\ref{k_CKY}). 
Tod has shown that the desired overall factor $z$ in the Przanowski-Tod metric (\ref{PT})
takes the following form\footnote{One can also show that the property  
$z\propto W^{-1}$ holds, where $W=\sqrt{P^{(i)}P^{(i)}} $ defines the superpotential and 
$P^{(i)}$ are the triplet of moment maps in quaternionic manifolds satisfying 
$J^{(i)}{}_{\mu\nu}U^\nu=\partial_\mu P^{(i)}+\epsilon_{ijk}\mas A^{(j)}{}_\mu P^{(k)}$. 
}
\begin{align}
\label{zsq}
z^{-2}=  k_{\mu\nu}[U]k^{\mu\nu}[U] \,. 
\end{align}
For any choice of parameters ($c_1, c_2$), the metric can be locally cast into the form (\ref{PT}) and 
one can also check that Toda equation is satisfied.\footnote{Since the Toda equation can be written into a covariant three-dimensional harmonic equation on the base space, there is no need to introduce the coordinate $x$ explicitly for this purpose.} However, this argument does not make any reference to the global issue. Here,  we wish to find the explicit coordinate transformation from (\ref{PD}) to (\ref{PT}) in order to see rigorously the existence of Killing spinors.

A key observation for this purpose is the fact that the PD family is conformal to K\"ahler metrics (\ref{Kahler_conformalCarter}), and the Przanowski-Tod metric is conformal to the Lebrun metric (\ref{LeBrun}) which is K\"ahler with an anti-self-dual K\"ahler form (\ref{KF_LeBrun}). Indeed, the identification of (\ref{Kahler_conformalCarter}) with the Lebrun metric (\ref{LeBrun}) turns out true. 
Accordingly, we get  
\begin{align}
\label{zsol_PD}
z=-\frac{3}{\Lambda}\frac{1+bp q}{p+q}\,. 
\end{align}
It is easy to check that the above $z$ indeed solves the conformally coupled scalar equation (\ref{CCeq}), and  corresponds to the choice $c_1=-1, c_2=-b$. 
This solution is beyond the separable ansatz discussed in ref.~\cite{Nozawa:2008wf} 
for conformal wave equations in PD family. Alternatively, the conformal factor can be 
determined by the relation 
$\hat g_{\mu\nu}=(C_{\rho\sigma\tau\lambda}C^{\rho\sigma\tau\lambda})^{1/3} g_{\mu\nu}$~\cite{Apostolov:2013oza,Derdzinski}, on the basis of ambi-K\"ahler structure.  
Also, it is not difficult to show that 
the metric  (\ref{Kahler_conformalCarter}) solves the Einstein-Maxwell equations (\ref{LeBrun_Ein}) and (\ref{LeBrun_Max}), by taking the structure functions of (\ref{Kahler_conformalCarter}) as  (\ref{PQ_fun}) with the self-dual condition (\ref{SD_W}). The geometry of this LeBrun metric is explored in 
appendix~\ref{sec:typeD_LeBrun}. 

Since the candidates for the coordinate $z$ and the supersymmetric Killing field $U$ are specified, we can  then show all the bilinear equations following the same procedure as the Lorentzian case~\cite{KN}.  
By the coordinate transformation
\begin{align}
\label{tyx_rel1}
t= -\frac{\tau}2 -\frac{\sigma}{2b}\,, \qquad
y=-\frac{\tau}2+\frac{\sigma}{2b}
\,, \qquad 
\D x = \frac{1-bq^2}{2bQ(q)}\D q+ \frac{1-bp^2}{2bP(p)}\D p  \,, 
\end{align}
we can see that the self-dual PD solution can be brought to the 
Przanowski?Tod form (\ref{PT}) with
\begin{subequations}
\label{PT_pq}
\begin{align}
H=&-\frac{\Lambda(p-q)(p+q)^3}{3[(1-bq^2)^2P(p)-(1-bp^2)^2Q(q)] }
\,, \\ 
\omega =&\frac{(1-b^2q^4)P(p)-(1-b^2p^4)Q(q)}{(1-bq^2)^2P(p)-(1-bp^2)^2Q(q)}\D y \,, \\ 
e^\varphi =&-\frac{36b^2 P(p)Q(q)}{(p+q)^4 \Lambda^2} \,.
\end{align}
\end{subequations}
One can verify that these quantities satisfy the continuous Toda equation (\ref{Toda}) and (\ref{PT_omega}). 
Up to this point, all equations are valid for both sign of the cosmological constant. 
In the $\Lambda<0$ case, one can reconstruct the bilinear tensors (\ref{bilinears}) out of 
$H$ and $\varphi$, and are given by  
\begin{subequations}
\begin{align}
\label{}
E=&\frac{(p-q)(p+q)^3+\ell^2 [(1-bq^2)^2 P(p)-(1-bp^2)^2 Q(q)]}{2\ell (p-q)(p+q)^2(1+b pq)}\,,\\ 
B=&\frac{(p-q)(p+q)^3-\ell^2 [(1-bq^2)^2 P(p)-(1-bp^2)^2 Q(q)]}{2\ell (p-q)(p+q)^2(1+b pq)}\,,\\
U=& -\partial/\partial\tau -b \partial/\partial\sigma \,, \\
V=&\D \left(-\frac{p+q}{1+b p q}\right) \,. 
\end{align}
\end{subequations}
In this case, all the bilinear equations are satisfied 
by the self-dual two-form 
\begin{align}
\label{F_KSpq}
F_{\rm BPS}=& \frac {\ell}{(p-q)(p+q)^3}\left[-P(p) \D q \we (\D \tau-q^2 \D \sigma)+Q(q) 
\D p \we (\D \tau -p^2 \D \sigma)
\right.\nonumber \\ 
&\left.  
+F_1\D q \we (\D \tau-p^2 \D \sigma)
-F_1\D p \we(\D \tau -q^2 \D \sigma)
\right] \,,
\end{align}
where $F_{\rm BPS}=\star F_{\rm BPS}$, $\D F_{\rm BPS}=0$ and 
\begin{align}
\label{F1}
F_1=\frac{p^2 q^2}{\ell^2}+(a^2-m^2) (1+b^2p^2q^2)-(p^2+q^2)\left(\frac 12+2m\frac{1-bpq}{p-q} \right)\,.  
\end{align}
We have now proved that the self-dual PD metric preserves at least 1/4 of supersymmetries.  
Note that the self-dual two-form (\ref{F_KSpq}) is essential for obtaining the Killing spinor, 
although it fails to contribute to the Einstein equations. 

It is worth emphasizing that 
the two-form (\ref{F_KSpq}) is not the one obtained by the self-dual limit of the original Maxwell field (\ref{PD_A}) for the PD metric. 
This is not surprising because the bilinears for the non-self-dual case (\ref{bilinears_nonSDPD}) are ill-defined in the self-dual limit. 
The above self-dual two-form $F_{\rm BPS}$ is not proportional to $J_+$. Instead, it can be written as a linear combination of $J^{(1)}_+\equiv J_+$ and 
$J^{(2)}_+\equiv e^1\we e^3+e^4\we e^2$ with ($q,p$)-dependent coefficients.  
One may then wonder where this type of self-dual harmonic two-form (\ref{F_KSpq}) originates from. 
A decisive clue again comes from the LeBrun metric, which solves the Einstein-Maxwell system (\ref{LeBrun_Ein})--(\ref{LeBrun_Max}).  Since the Maxwell equation is conformally invariant, 
the Maxwell field (\ref{LeBrun_Max}) of the LeBrun space is also a harmonic 2-form for the self-dual PD metric. One sees that the self-dual part of the two-form (\ref{LeBrun_Max}), corresponding to the Ricci-form of LeBrun, gives rise to the above two-form (\ref{F_KSpq}) as
\begin{align}
\label{}
F_{\rm BPS}= \frac{\ell (1+b pq)^2}{(p-q)(p+q)^3}\left(-F_1 J^{(1)}_+ +\sqrt{-P(p)Q(q)} J^{(2)}_+\right) =-\frac{\ell}2 \mathfrak R\,,
\end{align}
where the explicit Ricci form $\mathfrak R$ can be found in (\ref{RF}). 
Indeed, one can verify from (\ref{Asol}), (\ref{CT}), 
(\ref{PT_omega_app}), (\ref{LeBrun})  
that $F_{\rm BPS}=-(\ell/2)\mathfrak R$ is universally true for self-dual BPS solutions 
(see \cite{Farquet:2014kma}).

Let us next look for the second independent Killing spinor. In the original coordinates ($q,p $), the function $z$ for the desired transformation fails to be a meromorphic function of ($q, p$), so that it is awkward to obtain the Przanowski-Tod metric explicitly. We can avoid this difficulty by simply working in the 
($u, v$) coordinates.  As we discussed at (\ref{U_c1c2}), we need to obtain the vector field of the form $U=c'_1 \partial/\partial{\psi} +c'_2 \partial/\partial{\chi}$ in the ($u, v$) coordinate system, 
for which the appropriate conformal factor $z$ of Przanowski-Tod form is determined via (\ref{zsq}). 

Before proceeding, let us pause a little bit here and remark the convention that we employ. 
The previous section has shown that the transformation to the Carter form reverses the orientation, i.e, 
$e^1\we e^2\we e^3\we e^4 =-e^{(1)}\we e^{(2)}\we e^{(3)}\we e^{(4)}$. We have, however, 
fixed the orientation in appendix~\ref{app:SUSY_PD} to $\epsilon_{1234}>0$ 
for obtaining the supersymmetric canonical form. Thus,  we now flip the orientation  of (\ref{Carter_frame}) in such a way that  $e^{(1)}\we e^{(2)}\we e^{(3)}\we e^{(4)}$
is positively-oriented, for which the Weyl curvature of the Carter solution (\ref{Carter})
is {\it anti-self-dual}. 

With this convention in mind, we wish to identify the anti-self Carter metric~(\ref{Carter}), which is conformal to the K\"ahler metric (\ref{Kahler_conformalCarter}), as the LeBrun space. Then, the LeBrun metric must admit a self-dual K\"ahler form and the suitable form of $z$ can be deduced to be 
\begin{align}
\label{z_Carter}
z=-\frac{3}{\Lambda}\frac{1}{u+v} \,,
\end{align}
which corresponds to  $c'_1=1$ and $c'_2=0$. 
The relation (\ref{tausigma_tr}) reveals that this choice amounts to 
$U=\partial/\partial\tau -b\partial/\partial\sigma$ in the original coordinate
(modulo the scaling), which is obviously independent of the previous choice of the Killing vector 
$U=-\partial/\partial\tau-b \partial /\partial\sigma$. 
With this choice made,  the anti-self-dual Carter metric (\ref{Carter}) can be cast into the Przanowski-Tod form (\ref{PT}) by (cf. \cite{Farquet:2014kma})
\begin{subequations}
\label{PT_Carter}
\begin{align}
 H   &=-\frac{\Lambda}3\frac{(u-v)(u+v)^3}{\ma Q(u)-\ma P(v)} \,,  \\
 \omega   &= \frac{u^2 \ma P(v)-v^2 \ma Q(u)}{\ma Q(u)-\ma P(v)}\D y\,, \\
 e^\varphi&= -\frac{\ma Q(u)\ma P(v)}{(\Lambda/3)^2 (u+v)^4 } \,.   
\end{align}
\end{subequations}
with 
\begin{align}
\label{Carter_tyx}
t=\psi \,, \qquad y=\chi\,, \qquad 
\D x=\frac{\D u}{\ma Q(u)}+\frac{\D v}{\ma P(v)}\,. 
\end{align}
In the $\Lambda<0$ case, 
the bilinears are given by 
\begin{subequations}
\begin{align}
\label{}
E&=\frac{(u-v)(u+v)^3+\ell^2 [\ma Q(u)-\ma P(v)]}{2\ell (u-v)(u+v)^2} \,, \\
B&=-\frac{(u-v)(u+v)^3-\ell^2 [\ma Q(u)-\ma P(v)]}{2\ell (u-v)(u+v)^2} \,, \\
U&=\partial/\partial\psi \,, \\ 
V&= -\D (u+v) \,. 
\end{align}
\end{subequations}
The Maxwell field appearing in the Killing spinor equation is then given by 
\begin{align}
\label{F_uv}
F_{\rm BPS}=\frac{\ell}{(u-v)(u+v)^3}& \left[
\ma P(v)\D u\we (\D \psi-u^2\D \chi)-\ma Q(u) \D v\we (\D \psi-v^2 \D \chi)
\right. \nonumber \\
& \left. -F_2 \D u\we (\D \psi -v^2 \D \chi) +F_2 \D v \we (\D \psi-u^2 \D \chi)
\right]
\,,
\end{align}
where 
\begin{align}
\label{}
F_2=\frac{u^2 v^2}{\ell^2}+\alpha_0+(u^2+v^2) \left(\frac 12 \alpha_2+\frac{2\alpha_1}{u-v}\right)\,.
\end{align}
Here $(\alpha_0, \alpha_1, \alpha_2)$ appear in the structure functions 
as (\ref{Carter_PQ}). As in the previous case, the two-form (\ref{F_uv}) is distinct from the anti-self-dual limit of (\ref{A_uv}) and corresponds to the Ricci-form of LeBrun metric. We have now proven that the self-dual PD solution preserves precisely one half of supersymmetry. 

The PD metric has been transformed in two different ways into the Przanowski-Tod forms, for which 
the existence of two independent Killing spinors is made manifest. In principle, it would be possible to proceed without specifying the coefficients in (\ref{U_c1c2}) to show this claim. In that case, however, it is a notoriously difficult problem to find the explicit expression for the coordinate $x$. Without finding $x$, the global existence of the Killing spinor is not ensured. 

Przanowski analyzed the self-dual Einstein metric with a Killing vector $\xi$ and demonstrated that the 
metic is governed by a single complex master equation~\cite{Przanowski:1991ru}. He argued that two different classes of coordinates can be introduced, depending on the form of the Killing vector.  Letting $(\zeta ^1, \zeta^2)$ denote complex coordinates of hermitian metric, these classes are distinguished according to
 \begin{align}
\label{classAB}
\textrm{Class A}: \quad U =i (\partial/\partial\zeta ^2 -\partial/\partial \bar \zeta^2) \,, \qquad 
\textrm{Class B}: \quad U =i (\partial/\partial\zeta ^1 -\partial/\partial \bar \zeta^1) \,, \qquad 
\end{align} 
Later on, Tod was able to derive a local form of metric (\ref{PT}) for class B by starting from the Przanowski master equation, while for class A he only showed  the local existence of coordinates that reduce Przanowski's master equation to the Toda system. The coordinate transformation (\ref{pq_uv}) that we found is precisely of this kind. One can verify that in the original PD coordinates ($\tau ,\sigma, p, q$), the Przanowski-Tod metric (\ref{PT_pq}) falls into class B,  whereas the metric (\ref{PT_Carter}) falls into class A from the viewpoint of original coordinates but class B for the ($\psi, u, v, \chi$) coordinates, and vice versa. This change of classes traces back to the property (\ref{frame_rel}), where the orientation of the frame $e^a$ and $e^{(i)}$ are reversed.

\subsection{Transformations to the Calderbank-Pedersen metric}
\label{sec:CP}

In this subsection, we discuss some mathematical issues on the self-dual PD metric 
in the Przanowski-Tod form. See \cite{Farquet:2014kma} for a related discussion. 

Let us consider the three-dimensional base space 
$h_{mn}\D x^m \D x^n=\D z^2 +e^\varphi (\D x^2+\D y^2) $ of the Przanowski-Tod metric, 
which is conformal to the quotient of the manifold with the 
orbits of the bilinear Killing field $U^\mu=i \epsilon^\dagger\gamma_5\gamma^\mu \epsilon$. 
In the original coordinates, the line element reads  
\begin{align}
\label{base}
h_{mn}\D x^m \D x^n=\left(\frac{3}{\Lambda}\right)^2\left[-\frac{4b^2P(p)Q(q)}{(p+q)^4}\D y^2
+\frac{(1-bq^2)^2P(p)-(1-bp^2)^2Q(q)}{(p+q)^4}\left(
\frac{\D p^2}{P(p)}-\frac{\D q^2}{Q(q)}\right) \right]\,. 
\end{align}
As shown by Ward~\cite{Ward:1990qt}, this three-dimensional base space 
together with the continuous Toda equation (\ref{Toda}) defines the (representative metric of) Einstein-Weyl geometry \cite{GT,PT}. Namely, $\theta=\partial_z \varphi \D z$ satisfies 
$\ma D_m h_{np}=2 \theta_m h_{np}$, where $\ma D_m $ is a torsion-free 
affine connection called a Weyl connection \cite{GT,PT}. 
This is the Jones-Tod correspondence~\cite{Jones:1985}, 
relating the self-dual space and the Einstein-Weyl space with a Toda structure. 
Since $\theta_m $ fails to be closed, the Einstein-Weyl space (\ref{base}) is not conformal to the Einstein space.  As far as the authors know, our metric (\ref{base}) provides a new example of the (possibly noncompact) Einstein-Weyl space, although the metric cannot be expressible in terms of ($x, z$). The other base space stemming from the Carter expression (\ref{z_Carter})--(\ref{Carter_tyx}) is not independent but is simply given by $b\to 0$ limit with suitable rescaling of $y$ coordinate. The Einstein-Weyl space forms the base space of M-theory supersymmetric solutions with ${\rm SO}(6)\times {\rm SO}(3)$ invariance~\cite{Lin:2004nb}.

Since the Toda equation is nonlinear, it is difficult to extract physical and geometric information 
from (\ref{PT_pq}) and (\ref{PT_Carter}). Nevertheless, the second independent Killing vector, 
which commutes with the one built out of a Killing spinor, linearizes the governing equation. 
Suppose $\partial/\partial y$ is a Killing vector independent of $U=\partial/\partial t$. 
As shown by Ward~\cite{Ward:1990qt}, there exists a B\"acklund transformation
\begin{align}
\label{}
z=\rho \partial_\rho V \,, \qquad x=\partial_\eta V \,, \qquad e^\varphi =\rho ^2 \,,
\end{align}
which brings the ${\rm SU}(\infty)$ Toda equation (\ref{Toda}) into three-dimensional 
axisymmetric Laplace equation
\begin{align}
 \frac{1}{\rho }\partial_\rho (\rho \partial_\rho V) +\partial_\eta ^2
 V=0 \,.
 \label{hol_Toda_Laplace}
\end{align}
Once $V=V(\rho, \eta)$ is given, all the metric components are derived successively. 
The most remarkable feature is the linearity of governing equation. Thus, the solution is arbitrarily superposed and one can construct lots of solutions by taking Weyl solutions in vacuum general relativity. 
However, since the metric components involve the second derivative of $V$, it is not a priori clear which choice of the harmonic function on $\mathbb R^3$ yields a desired solution.

Calderbank and Pedersen obtained an explicit Einstein metric with a self-dual Weyl curvature admitting two commuting Killing vectors~\cite{Calderbank:2001uz}, on the basis of Joyce's construction~\cite{Joyce} in which the (not necessarily Einstein) spaces of self-dual Weyl curvature admitting surface-orthogonal ${\rm U}(1)^2$ actions are obtained. Their expressions are rather explicit and have more advantages over the Ward form (cf. \cite{Santillan:2004ef}). Equation (\ref{hol_Toda_Laplace}) implies the existence of a dual potential $\ti V$ such that 
\begin{align}
\label{}
\partial_\rho \ti V=\rho \partial_\eta V\,, \qquad 
\partial_\eta \ti V=-\rho \partial_\rho V \,,
\end{align}
satisfying $(\partial_\rho^2+\partial_\eta ^2)\ti V=\frac{1}{\rho}\partial_\rho \ti V$. 
Let us further define a function $G=G(\rho , \eta)$ by $G\equiv \rho^{-1/2}\partial_\eta \ti V$, 
which obeys a differential equation 
\begin{align}
\label{}
(\partial_\rho^2 +\partial_\eta ^2)G=\frac{3}{4\rho^2}G \,.
\end{align}
Namely, $G$ is an eigenfunction of the Laplacian on two-dimensional 
hyperbolic space $\rho^{-2}(\D \rho^2+\D \eta^2)$ with an eigenvalue $3/4$. 
It follows that the metric can be written as 
\begin{align}
\ell^{-2}\D s^2=&\frac{(G+2\rho G_\rho )^2+4\rho^2 G_\eta ^2}{\rho G^2[4\rho^2(G_\rho^2+G_\eta^2)-G^2]}\left[\D t\mp 
\left(\eta -
\frac{4\rho^2 G_\eta G }{(G+2\rho G_\rho )^2+4\rho^2 G_\eta ^2}\right)\D y\right]^2
\notag\\&
+\frac{\rho [4\rho^2(G_\rho^2+G_\eta^2)-G^2]}{G^2[(G+2\rho G_\rho )^2+4\rho^2 G_\eta ^2]}\D y^2
+\frac{4\rho^2(G_\rho^2+G_\eta ^2)-G^2}{4\rho^2 G^2}(\D \rho^2+\D \eta ^2)\,,
\label{CP}
\end{align}
where the subscript denotes the differentiation with respect to the argument and the upper (lower)
sign corresponds to the self-dual (anti-self-dual) case.  
The Calderbank-Pedersen metric appears as the moduli space of the universal hypermultiplet
with 1-loop string corrections \cite{Antoniadis:2003sw}. It is noteworthy that the self-dual PD family is hermitian, 
whereas the general Calderbank-Pedersen metric is not necessarily hermitian. 

We are now in a position to explore the Calderbank-Pedersen master variable $G$ in 
two different ways. It is not difficult to show that with the choice
\begin{subequations}
\begin{align}
\label{}
\rho &=\frac{2b \ell ^2\sqrt{-P(p)Q(q)}}{(p+q)^2} \,,  \\
\eta &=-\frac{2b\ell^2}{(p+q)^2}\left[\frac{p^2q^2 }{\ell^2}
+p q+(a^2-m^2)(1+b p^2q^2)-m(p-q)(1-bpq) 
\right] -[1+2b^2\ell^2(a^2-m^2)]\,, 
\end{align}
\end{subequations}
and 
\begin{align}
\label{}
G=z\rho^{-1/2}=\frac{\ell}{\sqrt{2b}}\frac{1+b p q}{[-P(p)Q(q)]^{1/4}},
\end{align}
the Calderbank-Pedersen metric (\ref{CP}) gives the self-dual PD solution. 

Working with the ($u, v$) coordinates, 
one can obtain another form of the Calderbank-Pedersen metric by
\begin{align}
\label{}
\rho=\frac{\ell ^2\sqrt{-\ma Q(u)\ma P(v)}}{(u+v)^2}\,, \qquad 
\eta =-\frac{\ell^2}{(u+v)^2}\left(\frac{u^2v^2}{\ell^2}+\alpha_0+\alpha_1(u-v)-\alpha_2 uv\right)\,,
\end{align}
with 
\begin{align}
\label{}
G=\frac{\ell}{[-\ma P(v)\ma Q(u)]^{1/4}} \,. 
\end{align}
Since these are very complicated functions in the original ($q,p$) coordinates, 
it is unlikely to find them explicitly. The derivation here considerably relies 
on the discovery of new coordinates ($u, v$) in the self-dual Carter form.  
However, it seems notoriously difficult to express the master variable $G$
in terms of ($\rho , \eta$) in an obvious fashion.


\subsection{Twistor equation}
\label{sec:twistor}

The twistor is a significant notion closely related to the Einstein space with a self-dual curvature. 
When the Maxwell field satisfies the self-dual condition, 
we found two independent Killing spinors $\epsilon$ satisfying (\ref{KS_AdS}). 
We now discuss the relationships of the Killing spinors and the twistor spinors. 

In terms of the solution $\epsilon$ of a Killing spinor (\ref{KS_AdS}), 
let us define the chiral spinors 
\begin{align}
\label{twistorspinor}
\zeta^\pm\equiv \frac 12(1\pm \gamma_5)\epsilon \,. 
\end{align}
After some manipulations, one sees that these spinors satisfy 
\begin{align}
\label{}
0=\left(\nabla_\mu -\frac i\ell A_\mu\right) \zeta ^\pm +\frac 1{2\ell} \gamma_\mu \zeta ^\mp \,. 
\end{align}
It therefore follows that these spinors obey the charged twistor equations~\cite{Semmelmann}
\begin{align}
\label{}
0=D_\mu \zeta^\pm -\frac 14 \gamma_\mu \gamma^\nu D_\nu \zeta^\pm \,, \qquad 
D_\mu \zeta^\pm \equiv \left(\nabla_\mu -\frac i\ell A_\mu \right)\zeta ^\pm \,. 
\end{align}
where $D_\mu $ is the ${\rm U}(1)$ gauge covariant derivative.  
Twistor equations imply the existence  CKY tensors as bilinears
\begin{align}
\label{}
k _{\mu\nu}^\pm =(\zeta^\mp) ^\dagger \gamma_{\mu\nu}\zeta^\mp \,,
\end{align}
which are (anti-)self-dual $\star k^\pm =k^\pm$ and satisfy the CKY equation~(\ref{CKYeq}). 
This is precisely the one constructed out of the Killing vector as (\ref{k_CKY}). 
In the earlier work \cite{Tod:2006wj}, only the neutral twistor equations have been worked out. 
Unfortunately, these neutral twistors are not directly built out of the Killing spinors in Einstein-Maxwell-$\Lambda$ system, as we have illustrated in (\ref{twistorspinor}). 
This means that in the Einstein manifolds with a self-dual Weyl curvature, the supersymmetry and the hidden symmetry are closely related.

\section{Hyper-K\"ahler metric}
\label{sec:HK}

In the Ricci-flat case, the self-duality of the Weyl tensor reduces to the 
self-duality of the Riemann tensor
\begin{align}
\label{}
R_{\mu\nu\rho\sigma} =\frac 12 \epsilon_{\mu\nu\tau\lambda}R^{\tau\lambda}{}_{\rho\sigma}
 \,. 
\end{align}
The four-dimensional manifold with a self-dual Riemann tensor is referred to as a hyper-K\"ahler space. 
This section discusses the transformation of the self-dual Ricci-flat PD family to the Gibbons-Hawking space in which the triholomorphic isometry is manifest, and the properties of gravitational instanton.

\subsection{Transformation to the Gibbons-Hawking metric}

The hyper-K\"ahler metric admitting a (triholomorphic) symmetry $U=\partial/\partial t$ is described by the Gibbons-Hawking space~\cite{Gibbons:1979zt}
\begin{align}
\label{GH}
\D s^2= H^{-1} (\D t + \omega )^2 + H \D {\bf x}^2 \,. 
\end{align}
where in the 3-dimensional vector notation the metric components 
obey the following linear system 
\begin{align}
\label{GH_Homegaeq}
\vec \nabla ^2 H=0 \,, \qquad \vec \nabla \times \vec \omega= \vec \nabla H \,.
\end{align}
$\vec \nabla$ is the ordinary derivative operator in Cartesian coordinates. 

We can show that the PD metric with a self-dual Riemann tensor belongs to the Gibbons-Hawking space. We do not attempt to repeat the detail, but show exclusively the final upshot. For $b\ne 0$, we define 
\begin{align}
\label{tytausigma}
t=\frac{\tau}{2}+\frac{\sigma}{2b} \,, \qquad y=\frac{\sigma}{2b}-\frac{\tau}2\,. 
\end{align}
The dimensional reduction of the metric (\ref{PD}) along a Killing vector 
$U=\partial/\partial\tau +b\partial/\partial\sigma =\partial/\partial t$ gives rise to 
the Gibbons-Hawking space (\ref{GH}) with 
\begin{align}
\label{hyperK_Homega}
H= \frac{(1+b p q)^2(p^2-q^2)}{(1-bq^2)^2 P(p)-(1-bp^2)^2 Q(q)}
\,,  \qquad 
\omega =
\frac{(-1+b^2q^4)P(p)+(1-b^2p^4)Q(q)}{(1-bq^2)^2 P(p)-(1-bp^2)^2 Q(q)}
 \D y\,, 
\end{align}
where the flat base metric  reads
\begin{align}
\label{flatbase}
\D {\bf x}^2 = -\frac{{4b^2}P(p)Q(q)}{(1+b p q)^4} \D y^2 +\frac{(1-bp^2)^2 Q(q)
-(1-bq^2)^2 P(p)}{(1+b p q)^4} \left(\frac{\D q^2}{Q(q)}-\frac{\D p^2}{P(p)}\right) \,.
\end{align}
The structure functions are given by 
(\ref{PQ_fun}) with $\Lambda =0$ and the self-duality restriction (\ref{SD_PD}).  
It is a simple exercise to check that (\ref{hyperK_Homega}) satisfies (\ref{GH_Homegaeq}). 
The base space can be transformed to the 
Cartesian coordinates $\D {\bf x}^2=\D x_1^2+\D x_2^2+\D x_3^2$ by
\begin{align}
\label{}
&x_1+i x_2 =\frac{\sqrt{-P(p)Q(q)}}{\gamma^{1/2}(1+b p q)^2} \exp({2ib} \gamma^{1/2} y)\,, \notag \\
&x_3  = \frac{b(m^2-a^2)(p^2+q^2)+m(p-q)(1-bpq)-pq}{\gamma^{1/2}(1+b pq)^2}  \,, 
\end{align}
with $\gamma=2b(m^2-a^2)^2+a^2$.

Taking the orthonormal frame 
\begin{align}
\label{}
\ti e^1=H^{-1/2}(\D t+\omega)\,, \qquad 
\ti e^{i+1}= H^{1/2}\D x^i \quad (i=1,2,3) \,, 
\end{align}
the hyper-complex structures $\ti J^{(i)}$ are given by
\begin{align}
\label{}
\ti J^{(i)}=\ti e^1 \we \ti e^{i+1} -\frac 12 \epsilon_{ijk} \ti e^{j+1} \we \ti e^{k+1} \,, \qquad 
\star \ti J^{(i)}= -\star \ti J^{(i)} \,, \qquad \nabla_\mu \ti J^{(i)} {}_{\nu\rho}=0
\end{align}
Note that these are distinct from those defined in (\ref{Ji_ASD}) for the quaternionic structure. 
Quantities defined in (\ref{Ji_ASD}) are not covariantly constant in the $\Lambda\to 0$ limit. 
In section~\ref{sec:Carter}, we have worked out the coordinate transformation of the self-dual PD metric to the Carter family. This is not applicable for $\Lambda=0$, in which the KY tensor~(\ref{KY_PD}) is degenerate and proportional to one of the hyper-K\"ahler forms 
\begin{align}
\label{}
f=-3b \gamma^{1/2} \ti J^{(3)} \,. 
\end{align}
Despite this fact, the hyper-K\"ahler PD metric is half-BPS, as shown in appendix~\ref{app:SUSY_PD}. 
In this case the self-dual Maxwell field appearing in the Killing spinor is given $F_{\rm BPS}=\frac 12 \D U$, 
where $U=H^{-1}(\D t+\omega)$ in the Gibbons-Hawking coordinates (\ref{GH}). 
The Killing vector plays the role of a vector potential for the source-free Maxwell equations in the Ricci-flat space. It is noteworthy to remark that this two-form $F_{\rm BPS}$ is not the self-dual limit of the original Maxwell field computed from (\ref{PD}). 


Since the PD metric with $\Lambda=0$ falls into the Weyl-Papapetrou form, 
any dimensional reduction along a Killing vector yields the Ernst system coupled to three dimensional gravity. In general, the scalar fields parameterizing the nonlinear sigma model do not decouple from the three-dimensional gravity, giving rise to a curved three-dimensional base space. The decoupling occurs precisely when we choose the Killing vector 
$U=\partial/\partial \tau+b \partial/\partial \sigma$ as in the above case, for which the base space is $\mathbb R^3$.

\subsection{Self-dual gravitational instanton}

Let us move on to the discussion for the gravitational instantons.  
Since the compact and regular hyper-K\"ahler manifolds are exhausted by $T^4$ or K3, 
we focus on the parameter range under which the manifold is noncompact. 

Since the $m=0$ case reduces to the flat space, we assume $m\ne 0$
in what follows. Let us define
\begin{align}
\label{}
R_\pm =\sqrt{x_1^2+x_2^2+(x_3- \beta_\pm )^2} \,, 
\qquad 
\beta_\pm =\frac{2b(m^2-a^2)-1\pm \beta^{1/2}}
{4b\gamma^{1/2}}  \,,
\end{align}
where $\beta=[2b(m^2-a^2)+1]^2-16bm^2$. 
It follows that 
\begin{align}
\label{Risol}
R_-= -\epsilon_-\gamma_- \frac{1-b pq+\delta_-(p-q)}{1+b p q} \,, \qquad 
R_+= \epsilon_+\gamma_+ \frac{1-b pq+\delta_+(p-q)}{1+b p q}\,, \qquad 
\end{align}
where 
\begin{align}
\label{}
\delta_\pm &=\frac{2b(m^2-a^2)+1\pm \beta^{1/2}}{4m} \,,\nonumber \\
\gamma_\pm &=\frac 1{4b\gamma^{1/2}} \left[(1-6b(m^2-a^2)\mp \beta^{1/2})
(1+2b(m^2-a^2)\mp \beta^{1/2})\right]^{1/2} \,.
\end{align}
$\epsilon_\pm =\{1,-1\}$ are chosen to ensure $R_\pm \ge 0$, 
depending on the coordinate ranges of ($q, p$). 
One finds that the metric functions given in (\ref{hyperK_Homega}) can be written as~\cite{Casteill:2001zk}
\begin{align}
\label{}
H=\frac{M_+ }{R_+}+\frac{M_- }{R_-}\,, \qquad 
\omega =\left(\frac{M_+(x_3-\beta_+)}{R_+}+\frac{M_-(x_3-\beta_-)}{R_-}\right) 
\frac{x_1\D x_2-x_2 \D x_1}{x_1^2+x_2^2}\,. 
\end{align}
where $M_\pm =\epsilon_\pm \beta^{-1/2}\gamma_\pm $. 
Therefore, the metric is asymptotically locally Euclidean~\cite{Eguchi:1980jx}.  
This is a generalization of the Eguchi-Hanson space 
$H=\frac{r_0}{R_+}+\frac{r_0}{R_-}$~\cite{Eguchi:1978xp} into distinct point sources. 
However, the two-center solution with unequal masses suffers from the Dirac-Misner string~\cite{Gibbons:1979zt}. Therefore, the regularity imposes $M_+=M_-\equiv M$, which is realized when 
\begin{align}
\label{}
m=\pm \sqrt{a^2+\frac 1{2b}} \,, \qquad 
{\rm or} \qquad m= \frac{1\pm \sqrt{a^2b+1/2}}{\sqrt b} \,.
\end{align}
Since the latter corresponds to $\beta =0$, we choose the former. 
This amounts to $\beta_+=-\beta_-$, i.e., 
we have now left a two-parameter family of solution characterized by ($M, \beta_+$). 
Even so, one of the parameters is gauged away,  
reducing to the Eguchi-Hanson space. To confirm this, let us  
move on to a new coordinate system defined by
\begin{align}
\label{}
x_1+i x_2=\frac 14 \sqrt{r^4-r_0^4}\sin \theta e^{i\phi} \,, \qquad 
x_3=\frac 14 r^2 \cos\theta \,, \qquad \beta_+ =-\frac{r_0^2}4 \,, \qquad 
\psi=\frac{t}{2M} \,.
\end{align}
We then get 
\begin{align}
\label{}
\D s^2=2 M \left[\left(1-\frac{r_0^4}{r^4}\right)\frac{r^2}4(\D \phi+\cos\theta \D \psi)^2+\frac{\D r^2}{1-r_0^4/r^4}+\frac{r^2}4(\D \theta^2+\sin^2\theta \D \psi^2 )\right] \,, 
\end{align}
as we desired. Therefore, the regular $b\ne 0$ PD metric with a self-dual Riemann tensor is exhausted by the Eguchi-Hanson space.

In the $b\to 0$ limit, we get the Euclidean Kerr-NUT solution, whose non-self-dual version was originally analyzed in \cite{Gibbons:1979nf}. The regularity of this class of metric has been revisited in some recent papers~\cite{Ghezelbash:2007kw,Chen:2010zu}. In particular, the Kerr-NUT metric with a self-dual Riemann tensor is isometric to the Taub-NUT solution~\cite{Chen:2010zu}. It therefore follows that the `acceleration parameter' $b$ controls the asymptotic behavior of instantons, viz, 
asymptotically locally Euclidean for $b\ne 0$ and asymptotically locally flat for $b=0$. 

As we have shown, the PD family with a self-dual Riemann tensor involves only the known gravitational instantons. It is an interesting future work to see if the non-self-dual PD family would be complete and non-singular.



\section{Summary and final remarks}
\label{sec:conclusion}

In this paper, we have elucidated a deep connection between the hidden symmetry and the supersymmetry 
for the Riemannian manifolds with self-dual curvatures in Einstein-Maxwell-$\Lambda $ system. We made a comprehensive analysis on the self-dual PD metric. We worked out some open issues in the literature and our results would establish a solid base for further investigations of quaternion K\"ahler geometries. 

We found that the Euclidean PD metric admits two integrable complex structures with different orientations, whose intrinsic torsions are expressed in terms of the locally exact Lee forms. This permits us to perform the conformal transformations to the toric K\"ahler metrics (\ref{Kahler_conformalCarter}), 
consistent with the results in~\cite{Apostolov:2013oza}.
These K\"ahler spaces may be instrumental as the base space for the construction of black holes in five-dimensional gauged supergravity~\cite{Maeda:2011sh}. 

We found that the nondegenerate rank-two KY tensor for the self-dual PD space found in \cite{Houri:2014hma} stems naturally from the linear combinations of the CKY tensor (\ref{CKY_Y}) characterizing the type-D property and the CKY tensor (\ref{k_CKY}) characterizing the Einstein manifolds with a self-dual Weyl tensor. Section~\ref{sec:twistor} demonstrated that the latter CKY tensor is profoundly linked to the charged twistor spinor. 
According to the general proof given in \cite{Houri:2007xz}, the existence of a nondegenerate rank-two KY tensor implies that the metric must belong to the (generalized) Carter family. This means that the self-dual PD metric must have a description in the Carter form. Taking the eigenvalues of the KY tensor as new coordinates, we found a coordinate transformation that casts the self-dual PD metric into the self-dual Carter metric with reversed orientation.  Namely, the acceleration parameter that distinguishes the PD family from the Carter family is gauged away and does not have an invariant meaning in the self-dual case. 

Using the two distinct fashions to describe the self-dual PD family, we next discussed the supersymmetry of the self-dual PD solution. While the first integrability condition is always satisfied in the self-dual case (\ref{KS_SD_int}), the construction of explicit Killing spinor is generally a hard task. Nevertheless, we were able to construct two independent Killing spinors for the self-dual case by rewriting the metric into two Przanowski-Tod forms. A crucial observation for the construction of a Killing spinor is that the Maxwell field 
is not given by the self-dual limit of the original gauge field, but is given by the Ricci form of the LeBrun metric. 

The base space of the Przanowski-Tod metric provides a new example of Einstein-Weyl space, which will be useful for constructing the space of reduced holonomy.
We were also able to translate the metric into two Calderbank-Pedersen forms. In particular, we found a Calderbank-Pedersen master variable $G$ that determines the metric completely for each case.  This description is more suited than the ${\rm SU}(\infty)$ Toda form, since the metric components are obtained immediately form $G$ and its first derivatives. 

Our constructions of quaternionic spaces have a number of applications. 
Following the arguments in \cite{Behrndt:2002xm}, one can find the type IIA solutions. 
It seems interesting to look for string vacua by gauging the two ${\rm U}(1)$ isometries of hypermultiplets in the context of $N=2$ gauged supergravities. Also, the M-theory embedding \cite{Lin:2004nb} using  our Einstein-Weyl space (\ref{base}) may appeal to string theorists. The two Przanowski-Tod metrics we found belong to the class A and B according to the classification of the Killing vector with respect to the complex structure (\ref{classAB}). In the context of instanton corrections to the universal hypermultiplet,  the class A appears for the fivebrane instantons, whereas the class B for the membrane instantons~\cite{Alexandrov:2006hx}. It is also intriguing to explore along this line.

In this paper we did not discuss the Euclidean supersymmetric solutions with $\Lambda>0$. 
For the Lorentzian signature, the positive cosmological constant is forbidden because it is not 
compatible with supersymmetry. Instead, it corresponds to the `fake supergravity.'
The fake supersymmetric solutions have focused attention recently,  
since they contain black holes in an expanding universe \cite{Nozawa:2010zg,Chimento:2012mg,Chimento:2014afa}. In Euclidean signature, there seems to be no consistent reason to discard the  $\Lambda>0$ case. In this case, the vector fields constructed as bilinears of Killing spinor fail to yield any Killing vectors. This is a technical obstruction to explicitly construct a Killing spinor by starting with (\ref{PD}). Perhaps, the Einstein-Weyl structure in the $\Lambda<0$ case discussed in section \ref{sec:CP} may potentially give some insights to this aim.  We hope to revisit this issue in the near future.


\section*{Acknowledgements} 
MN would like to thank Dietmar Klemm for stimulating discussions. 
The work of MN is partly supported by JSPS and INFN. 
TH would like to thank Maciej Dunajski for fruitful comments.
The work of TH is supported by the JSPS Grant-in-Aid for Scientific Research No.\ 26$\cdot$1237.

\appendix


\section{Classification of supersymmetric solutions} 
\label{app:SUSY_PD}

Let us review the classification scheme of supersymmetric backgrounds by means of 
spinor bilinears (cf. \cite{Caldarelli:2003pb,Klemm:2015mga}).  An alternative method exploiting the 
spinorial geometry can be found in refs. \cite{Dunajski:2010zp,Gutowski:2010zs,Dunajski:2010uv}.

\subsection{Canonical form of the metric for $\Lambda<0$}

As in the case of Lorentzian signature, 
the bilinears built out of the Killing spinor play an important role to classify the supersymmetric solutions. 
Given a Dirac spinor $\epsilon$, one can define the following tensorial quantities 
\begin{align}
 E\equiv \epsilon^\dagger \epsilon \,, \qquad B\equiv \epsilon^\dagger \gamma_5
 \epsilon \,, \qquad V_\mu \equiv \epsilon^\dagger \gamma_\mu \epsilon
 \,,\qquad 
U_\mu \equiv {i}\epsilon ^\dagger \gamma_5 \gamma_\mu \epsilon \,, \qquad 
\Phi_{\mu\nu }\equiv { i}\epsilon^\dagger \gamma_{\mu\nu }\epsilon \,,
\label{bilinears}
\end{align}
where $\gamma_5=\gamma_{1234}$ is a chiral matrix with $\gamma_5^\dagger =\gamma_5$. 
In this appendix, the volume form is fixed to have $\epsilon_{1234}=1$.  
With this convention, the above bilinear tensors are all real. 
It is also convenient to introduce supplementary complex tensors 
\begin{align}
\label{}
W_\mu \equiv \epsilon^TC^{-1}\gamma_\mu \epsilon \,, \qquad 
\Psi_{\mu\nu}\equiv i \epsilon^TC^{-1}\gamma_{\mu\nu} \epsilon \,,
\end{align}
where $C$ is the charge conjugation matrix satisfying 
$C^{-1}\gamma_\mu C =-\gamma_\mu^T$, $C^T=-C$. 
These bilinears obey algebraic and differential relations, and define the $G$-structures
which severely constrain the geometry and the flux. The algebraic relations arise from 
the Fierz identities and some prime examples are 
\begin{align}
\label{}
& V^\mu \gamma_\mu \epsilon=i U^\mu \gamma_5 \gamma_\mu \epsilon=(E- \gamma_5 B)\epsilon\,, 
\qquad  W^\mu \gamma_\mu \epsilon=0\,, \qquad V\cdot U=W\cdot W =V\cdot W=U\cdot W=0 \,, \notag \\
    & V^\mu V_\mu =U^\mu U_\mu =\frac 12 W^\mu \bar W_\mu = E^2- B^2>0  \,,  \qquad 
    (E^2-B^2) \Phi_{\mu\nu}= 2 BU_{[\mu }V_{\nu]}-E \epsilon_{\mu\nu\rho\sigma}U^\rho V^\sigma \,,
    \nonumber \\
   & (E^2-B^2)g_{\mu\nu}=V_\mu V_\nu+U_\mu U_\nu +W_{(\mu}\bar W_{\nu)}   \,, \qquad 
   (E^2-B^2)\Psi_{\mu\nu}=2 (BU_{[\mu}-i E V_{[\mu})W_{\nu]} \,.
\end{align}
It follows that two-forms $\Phi$ and $\Psi$ are redundant, and are expressible in terms of other bilinears. 
Assuming that  $\epsilon$ satisfies the Killing spinor equation for $\Lambda=-3\ell^{-2}<0$
\begin{align}
\label{KS_AdS}
 \hat \nabla_\mu \epsilon \equiv \left(\nabla_\mu +\frac{i}{4}F_{\nu\rho }\gamma^{\nu\rho }\gamma_\mu -
 \frac i\ell A_\mu +\frac{1}{2\ell}\gamma_\mu \right)\epsilon =0\,, 
\end{align}
the bilinears fulfill the following first-order differential relations
\begin{subequations}
\label{diff_eq}
\begin{align}
\nabla_\mu E & =-\frac{1}{\ell }V_\mu -\star F_{\mu\nu } U^\nu \,,\label{eq_E}\\
\nabla_\mu B& =F_{\mu\nu }U^\nu \,,\label{eq_B}\\
\nabla_\mu V_\nu &=-\frac{1}{\ell} g_{\mu\nu }E +2 F_{(\mu }{}^\rho
 \Phi_{\nu)\rho }-\frac{1}{2} g_{\mu\nu }F^{\rho\sigma }
 \Phi_{\rho\sigma } \,, \label{eq_V}\\
\nabla_\mu U_\nu &=-\frac{1}{\ell}\star \Phi_{\mu\nu }-E\star F_{\mu\nu }
-BF_{\mu\nu } \,, \label{eq_U}\\
\nabla_\mu \Phi_{\nu\rho } &=-\frac{1}{\ell } \epsilon_{\mu\nu\rho\sigma
 } U^\sigma +2F_{\mu[\nu }V_{\rho] }-V_\mu F_{\nu\rho } -2 g_{\mu [\nu } F_{\rho] \sigma
 }V^\sigma \,, 
\end{align}
\end{subequations}
with 
\begin{align}
 \nabla_\mu W_\nu = \frac{2i}{\ell }A_\mu W_\nu -\frac{i}{\ell}
\Psi_{\mu\nu }+2 F_{(\mu }{}^\rho \Psi_{\nu)\rho }-\frac{1}{2}g_{\mu\nu }F^{\rho\sigma }\Psi_{\rho\sigma } \,. 
\label{eq_W}
\end{align}
The above differential relations imply that $U$ is a Killing vector and $V$ is a
closed one-form. One can also check that $W$ is hypersurface-orthogonal 
$W \we \D W=0$. Since ($U_\mu, V_\mu, W_\mu, \bar W_\mu$) constitutes an 
orthonormal frame, 
it is then convenient to introduce the local coordinate system
\begin{align}
\label{BPSmetric_AdS}
 \D s^2 =f(\D t+\omega)^2 +f^{-1}  [ \D w^2 +e^{2\phi} (\D x^2+\D y^2)] \,, 
\end{align}
with 
\begin{align}
\label{W_phase}
U =\partial/\partial t \,, \qquad 
V =\D  w\,, \qquad W=e^{\phi}(\D x+i \D y) \,, \qquad 
f= E^2-B^2 \,. 
\end{align}
Here, $t $ is the Killing coordinate and the one-form $\omega$ describes the twist of $U$. 
The first two relations of (\ref{diff_eq}) can be solved to give the Maxwell field
\begin{align}
 F_{\mu\nu }=\frac{1}{f}\left[-2U_{[\mu }\nabla_{\nu
 ]}B+\epsilon_{\mu\nu\rho\sigma }U^\rho(\nabla^\sigma E+\ell^{-1} V^\sigma) \right] \,.
 \end{align}
 From the anti-symmetric part of (\ref{eq_W}), the  gauge potential $A_\mu $ reads 
\begin{align}
\label{Asol}
A=B (\D t +\omega)+\frac {\ell}2 (\partial_y \phi \D x-\partial_x \phi \D y) \,,
\end{align}
where we have employed a gauge $U^\mu A_\mu=B$ to ensure $\mas L_U A_\mu=0$. 
Requiring the Maxwell equation $\D \star F=0$, Bianchi identity $\D F=0$ and the
differential relation for $V$,  we can obtain the governing equations 
\begin{subequations}
\label{BPSsystem}
\begin{align}
 0&=\phi' +\frac{1}{2}(F_++F_-)\,, \\
0&=\Delta F_\pm +e^{2\phi}(F_\pm ^3-3 F_\pm F_\pm ' +F_\pm '') \,, \\
0&= \Delta \phi +\frac{1}{2}e^{2\phi}(-F_+'-F_-'+F_+^2+F_-^2-F_+F_-) \,, 
\end{align}
\end{subequations}
where the prime denotes the differentiation with respect to $w$ and 
\begin{align}
\label{}
F_\pm \equiv \frac{2}{\ell (E\pm B)} \,, \qquad \Delta=\partial_x^2+\partial_y^2 \,.
\end{align} 
The differential relation for $U$ (\ref{eq_U}) gives an 
equation for  $\omega$:
\begin{align}
\label{domega}
\D \omega = - \frac {1}{f^2} \star (U \we \Omega ) \,, \qquad 
\Omega \equiv 2 \left(B \D E-E \D B +\frac 2\ell B \D w \right) \,.
\end{align}
The integrability conditions of (\ref{domega}) are ensured by (\ref{BPSsystem}).

Taking the frame
\begin{align}
\label{}
e^1=f^{1/2}(\D t+\omega) \,, \qquad 
e^2=f^{-1/2}e^\phi \D x \,, \qquad e^3=f^{-1/2}e^\phi \D y \,, \qquad 
e^4= f^{-1/2} \D w \,, 
\end{align}
the Fierz identities imply
\begin{align}
\label{proj}
i \gamma^1\epsilon =f^{-1/2}(E\gamma_5-B) \epsilon \,,  
\qquad 
\gamma^{23}\epsilon =i \epsilon \,. 
\end{align} 
Working in a 
gauge $U^\mu A_\mu =B$ and imposing the first condition of (\ref{proj}), 
the Killing spinor is $t$-independent. 
Under the second condition of (\ref{proj}), the spatial components of the Killing spinor can be integrated 
as in the Lorentzian case and the solution is given by
\begin{align}
\label{sol_KS}
\epsilon = \frac 14\left(\sqrt{E+B}- i \gamma^1 \sqrt{E-B}\right) 
(1-i\gamma^{23})(1+\gamma_5) \epsilon_0 \,, 
\end{align}
where $\epsilon_0$ is a constant Dirac spinor and the 
phase of the spinor has been chosen appropriately. 

To summarize, the necessary and sufficient conditions for the preservation of supersymmetry 
in Euclidean Einstein-Maxwell-$\Lambda$ system is to solve the nonlinear system (\ref{BPSsystem}). In this case, 
 the solution preserves at least 1/4 of supersymmetry.

\subsubsection{(Anti-)self-dual solution} 

Let us next consider the case where  the Maxwell field is (anti-)self-dual $F=\pm \star F$.
 Together with (\ref{eq_E}) and (\ref{eq_B}), this implies 
\begin{align}
\label{EBrel}
E+\frac{w}\ell =\mp B \,. 
\end{align}
In this case, the BPS equations (\ref{BPSsystem}) are further simplified \cite{Dunajski:2010zp}. 
Defining 
\begin{align}
\label{CT}
z= -\frac{\ell^2}{w} \,, \qquad H^{-1} =\frac{2 E}{\ell }z -1 \,, \qquad 
e^\varphi=\frac{z^4}{\ell^4} e^{2\phi} \,, 
\end{align}
the solution can be cast into the 
Przanowski-Tod metric~\cite{Przanowski:1991ru,Tod:2006wj} 
\begin{align}
\label{PT_app}
\D s^2=\frac{\ell^2}{z^2} \left[
H^{-1} (\D t +\omega)^2 +H \{\D z^2+e^\varphi (\D x^2+\D y^2)\} 
\right] \,, 
\end{align}
where $ H$ and $\omega $ are  obtained by $\varphi$ as 
\begin{align}
\label{PT_omega_app}
\D \omega =\pm [\partial_x  H \D y \we \D z-\partial_y H \D x \we \D z +\partial_z (e^\varphi H)\D x \we \D y] \,, 
\qquad 
H=1 -\frac 12 z\partial_z \varphi \,.
\end{align}
$\varphi$ obeys the ${\rm SU}(\infty)$ Toda equation
\begin{align}
\label{Toda_app}
\Delta \varphi +\partial_z ^2 (e^\varphi)=0 \,,
\end{align}
where $\Delta=\partial_x^2+\partial_y^2$. 
The solution to the Killing spinor equation simplifies to 
\begin{align}
\label{KSsol_SD}
\epsilon =\frac 14 \sqrt{\frac \ell z}(
1\mp i H^{-1/2} \gamma^1)(1-i \gamma^{23})(1\pm \gamma_5)\epsilon_0 \,. 
\end{align}
Since the BPS metric with (anti-)self-dual Maxwell field necessarily takes 
the Przanowski-Tod form (\ref{PT_app}), it turns out that the Weyl tensor is also (anti-)self-dual.
This is a direct consequence of  the existence of a Killing spinor.

Przanowski \cite{Przanowski:1991ru}  and Tod \cite{Tod:2006wj} 
have shown that the Przanowski-Tod metric is hermitian. Here, 
it is interesting to observe that the Przanowski-Tod metric (\ref{PT_app}) is 
conformal to the LeBrun metric~\cite{LeBrun}, 
\begin{align}
\label{LeBrun}
\hat g_{\mu \nu}\D x^\mu \D x^\nu =
H^{-1} (\D t +\omega)^2 +H \{\D z^2+e^\varphi (\D x^2+\D y^2)\} \,,
\end{align}
where $H$, $\varphi$ and $\omega$ satisfy 
(\ref{PT_omega_app}) and (\ref{Toda_app}).\footnote{
More generally, $H$ is not directly related to $\varphi$ in the general LeBrun metric but
obeys the differential equation $\Delta H +\partial_z ^2 (e^\varphi H)=0$.
The restriction $H=1 -\frac 12 z\partial_z \varphi$ follows from the 
(anti-)self-duality. 
}
This is the K\"ahler metric with a vanishing scalar curvature. 
When the curvature is self-dual (resp. anti-self-dual), the K\"ahler form $\hat J$ is  anti-self-dual 
(self-dual)  $\hat J=\mp \hat \star \hat J $ 
and is given by 
\begin{align}
\label{KF_LeBrun}
 \hat J= (\D t +\omega )\we \D z \mp H e^\varphi \D x\we \D y \,,\qquad 
 \hat \nabla_\mu  \hat J_{\nu\rho} =0 \,, \qquad 
\mas L_{\partial/\partial t}  \hat J=0 \,. 
\end{align}
The LeBrun solution is realized as an exact solution to the Euclidean Einstein-Maxwell system
\begin{align}
\label{LeBrun_Ein}
\hat R_{\mu\nu} = 2 \left( \hat{\ma F}_{\mu\rho}\hat{\ma F}{}^{\nu\rho} -\frac 14
\hat g_{\mu\nu}\hat{\ma F}_{\rho\sigma}\hat{\ma F}{}^{\rho\sigma} \right) \,, \qquad 
\D \hat \star \hat{\ma F}=0\,, \qquad \D \hat{\ma F}= 0 \,,
\end{align}
 in which 
the Maxwell field $\hat{\ma F}$ is given by
\begin{align}
\label{LeBrun_Max}
\hat{\ma F}_{\mu\nu}= \frac{1}{\sqrt 2}\left( \hat J_{\mu\nu}+\frac{1}{2}\mathfrak R_{\mu\nu}\right)
 \,,
\end{align} 
where 
$\mathfrak R_{\mu\nu}\equiv\frac 12\hat R_{\mu\nu\rho\sigma}\hat J^{\rho\sigma}=\pm\hat\star \mathfrak R_{\mu\nu}$
is the Ricci-form of the LeBrun metric. 
If we choose the supersymmetric Killing field 
$U^\mu =i\epsilon^\dagger\gamma_5\gamma^\mu\epsilon$ to construct the anti-self-dual CKY 2-form (\ref{k_CKY}),  $h$ is conformal to the K\"ahler-form (\ref{KF_LeBrun}) on the LeBrun metric, 
which is a trivial KY tensor on the LeBrun.

\subsubsection{Supersymmetry of the non-self-dual Pleba\'nski-Demia\'nski family}

Here, let us discuss the supersymmetry of the non-self-dual PD metric to complete the research. 
The self-dual case has been argued in section \ref{sec:SUSY}. 

There exist a variety of different coordinates to describe the 7 parameter family of PD metric. 
In refs.~\cite{KN,Griffiths:2005qp}, the following coordinates ($\hat \tau, \hat \sigma, \hat p, \hat q$) were used to describe the PD family, 
\begin{align}
\D s^2 = &
\frac1{(1-\hat p\hat q)^2}\left[\frac{\hat Q(\hat q)}{\hat q^2-\hat\omega^2\hat p^2}(\D\hat \tau -\hat\omega \hat p^2\D\hat \sigma)^2 
+ \frac{\hat q^2-\hat\omega^2\hat p^2}{\hat Q(\hat q)}\D \hat q^2
\right.\nonumber \\ & \left.
+\frac{\hat q^2-\hat\omega^2\hat p^2}{\hat P(\hat p)}\D \hat p^2 
+ \frac{\hat P(\hat p)}{\hat q^2-\hat\omega^2\hat p^2}(-\hat\omega\D\hat \tau +\hat  q^2\D\hat \sigma)^2\right]\,, \label{metr-PD-eucl}\\
A=& -\frac{Q_m\hat p (-\hat\omega \D \hat \tau+\hat q^2\D \hat \sigma)
+Q_e \hat q(\D \hat \tau-\hat\omega \hat p^2 \D \hat \sigma)}{\hat q^2-\hat\omega^2 \hat p^2}\,, 
\end{align}
where the structure functions are given by 
\begin{align}
\hat P(\hat p) =& k + 2\hat\omega^{-1}n\hat p - \varepsilon \hat p^2 + 2m\hat p^3 + (\hat\omega^2 k - Q_m^2 +Q_e^2 + \hat\omega^2\Lambda/3)\hat p^4\,, \nonumber \\
\hat Q(\hat q) =& (-\hat\omega^2 k + Q_m^2 - Q_e^2) - 2m\hat q + \varepsilon \hat q^2 - 2\hat\omega^{-1}n\hat q^3 - (k + \Lambda/3)\hat q^4\,.
\end{align}
Since (\ref{metr-PD-eucl}) involves eight parameters 
($m, n , Q_e, Q_m, \Lambda, k, \varepsilon, \hat\omega$), 
one of the parameters are redundant and we can take it any values we wish as discussed in
\cite{Griffiths:2005qp}. 
Taking 
\begin{align}
\label{}
\varepsilon= -1 \,,
\end{align}
with the following redefinitions
\begin{align}
k& =(n^2-a^2) b^2 \,,  \qquad \hat\omega =1/b \,,\qquad 
\hat P(\hat p) =-b^2 P(p) \,, \qquad 
\hat Q(\hat q)=Q(q)\,, \notag \\
\hat \tau &=\tau \,, \qquad \hat \sigma =\sigma/b \,, \qquad \hat q= -q \,, \qquad 
\hat p =b p \,, 
\end{align}
one can recover the PD metric (\ref{PD}) employed in the body of text. 

In~\cite{KN}, the supersymmetry of the Lorentzian PD metric was addressed. 
To discuss the supersymmetry of the Euclidean non-self-dual PD solution, 
it is more convenient to work with the above coordinate system (\ref{metr-PD-eucl}), rather than
(\ref{PD}). 
The integrability conditions for the Killing spinor 
${\rm det}[\hat \nabla_\mu, \hat \nabla_\nu]=0$
give rise to the necessary conditions for supersymmetry and are 
obtained by the Wick rotation of the Lorentzian version
\begin{subequations}
\label{BPS}
\begin{align}
 &n [m^2-n^2-(Q_m^2-Q_e^2)\varepsilon ]+2 \hat\omega m (Q_e^2-Q_m^2)\left(\frac{\Lambda}{6}+k\right)\nonumber \\ & \qquad 
+\frac{\Lambda\hat\omega}{3}[2n Q_eQ_m-m(Q_m^2+Q_e^2)]=0 \,, \\
&(Q_m^2-Q_e^2)\left[m^2Q_m^2-n^2Q_e^2+(m^2-n^2)\left(\hat\omega^2k+\frac{\Lambda\hat\omega^2}{
6}-Q_m^2\right)\right]\nonumber \\ & \qquad 
+\frac{\Lambda\hat\omega^2}{3} 
\left[-2mnQ_eQ_m+\frac{1}{2}(Q_m^2+Q_e^2)(m^2+n^2)\right]=0\,.
\end{align}
\end{subequations}
These conditions are trivially satisfied for the (anti-)self-dual solutions (\ref{SD_PD}). 
This occurs for the general BPS solution due to the relation (\ref{KS_SD_int}). 

Assuming $\Lambda=-3\ell^{-2}<0$ and that 
the Weyl tensor and the Maxwell fields are not (anti-)self-dual,  
one can construct bilinears for Killing spinors by   
a suitable Wick-rotation of the ones given in~\cite{KN}. 
If $(Q_e^2-Q_m^2)(m^2-n^2)\ne 0$,  eq.~(\ref{BPS}) can be solved with respect to 
$k$ and $\varepsilon$ and 
it  is then straightforward to check that with 
\begin{subequations}
\label{bilinears_nonSDPD}
\begin{align}
 B&=-\frac{(-Q_e^2+Q_m^2)[ c_+(\hat q Q_e-\hat p Q_m \hat\omega)+c_-\hat p\hat q(\hat q Q_m-\hat p Q_e \hat\omega)]- c_+c_-(\hat q^2-\hat p^2 \hat\omega^2)}{(-Q_e^2+Q_m^2)(\hat q^2-\hat p^2 \hat\omega^2)} \,, \\
 E&=-\frac{ c_+^2-c_-^2 \hat p\hat q}{(1-\hat p\hat q)(-Q_e^2+Q_m^2)} \nonumber \\ 
& \quad +\frac{ c_+[Q_m (-\hat p^3 \hat\omega^3 +\hat q \hat\omega)+Q_e(\hat q^3-\hat p \hat\omega^2)]
-c_-[Q_m \hat q^2(-\hat p^3\hat\omega^2+\hat q)+Q_e\hat p^2 (-\hat p \hat\omega^3+\hat q^3 \hat\omega)]}{\hat\omega (1-\hat p\hat q)(\hat q^2-\hat p^2 \hat\omega^2)}\,, \\
U&= c_+\partial/\partial\tau +c_-\partial/\partial\sigma \,, \\
V&= \D \left[\ell\frac{(m^2-n^2)\hat\omega-(Q_m^2-Q_e^2)(m\hat p\hat\omega+n \hat q)}{\hat\omega(1-\hat p\hat  q)}\right] \,. 
\end{align}
\end{subequations}
where 
\begin{align}
\label{}
c_+=m Q_m-n Q_e\,, \qquad 
c_-=m Q_e-n Q_m\,, 
\end{align}
all the bilinear equations (\ref{diff_eq}) are satisfied. Equivalently, (\ref{sol_KS}) solves the 
Killing spinor equation.  The transformation to the canonical form (\ref{BPSmetric_AdS}) is given by 
\begin{align}
\label{}
t&=\frac{\tau}{2c_+}+\frac{\sigma}{2c_-}\,, \qquad 
y=\frac{\tau}{2c_+}-\frac{\sigma}{2c_-}\,, \qquad 
w=\ell\frac{(m^2-n^2)\hat\omega-(Q_m^2-Q_e^2)(m\hat p\hat\omega+n \hat q)}{\hat\omega(1-\hat p\hat  q)}\,,  \\
 \D x&=\frac{\ell[(Q_m^2-Q_e^2)(n+m\hat\omega \hat p^2)-(m^2-n^2)\hat p\hat\omega]}
 {2c_+c_-\hat\omega \hat P(\hat p)}\D \hat p
 -\frac{\ell[(Q_m^2-Q_e^2)(m\hat\omega+n \hat q^2)-(m^2-n^2)\hat q\hat\omega]}{2c_+c_-\hat\omega \hat Q(\hat q)}\D \hat q\,,\notag 
\end{align}
with
\begin{align}
\label{}
 \omega =\frac{(c_+^2-c_-^2\hat p^4\hat\omega^2)\hat Q(\hat q)+
 (c_+^2\hat\omega^2-c_-^2\hat q^4)\hat P(\hat p)}
 {(c_+-c_-\hat p^2\hat\omega)^2\hat Q(\hat q)+(c_+\hat\omega-c_-\hat q^2)^2\hat P(\hat p)}\D y
 \,, \qquad 
 e^{2\phi}=\frac{4c_+^2c_-^2\hat P(\hat p)\hat Q(\hat q)}{(1-\hat p\hat q)^4}\,.
\end{align}
This proves that the Euclidean non-self-dual PD metric admits a BPS limit. It is worth commenting that the non-self-dual metric only preserves one quarter of supersymmetry, since 
$[\hat \nabla_\mu, \hat \nabla_\nu]$ admits a single zero eigenvalue.  
In contrast, the self-dual metric preserves half of supersymmetry as we have shown in section \ref{sec:SUSY}.

\subsection{Canonical form of the metric for $\Lambda=0$}

It is instructive to review the $\Lambda=0$ case (see~\cite{Dunajski:2006vs} for the non-self-dual case).  
The differential relations for bilinears can be obtained by the $\ell\to \infty$
limit of (\ref{diff_eq}) and (\ref{eq_W}). 
Eq. (\ref{eq_W}) implies that $W$ can be expressed as $W=\D x+i \D y$. 
Thus, the metric can be written into (\ref{BPSmetric_AdS}) with $\phi=0$, 
for which $U=\partial/\partial t$ is a Killing vector and $V=\D w$. 
Hence, the three-dimensional base space is flat. 
The Maxwell equation and the Bianchi identity are combined to yield
$\vec \nabla^2 H_\pm =0$, where $H_\pm\equiv (E\pm B)^{-1}$. 
$\omega$ can be found by integrating (\ref{domega}). 
This is a Euclidean version of IWP solution~\cite{Perjes:1971gv,Israel:1972vx}. 

Using the first equation of projection conditions (\ref{proj}), 
the Killing spinor equation can be integrated to yield 
\begin{align}
\label{KSsol_Lambda0}
\epsilon = \frac 12\left(\sqrt{E+B}-i \gamma_1 \sqrt{E-B}\right) 
(1+\gamma_5) \epsilon_0 \,, 
\end{align}
where $\epsilon_0$ is a constant Dirac spinor. Since the Killing spinor 
(\ref{KSsol_Lambda0}) is subjected to the single projection 
$i \gamma^1\epsilon =f^{-1/2}(E\gamma_5-B) \epsilon $, 
the solution preserves half of supersymmetries. 

Let us next proceed to the case with a self-dual Maxwell field $F=\star F$. 
From the differential relation (\ref{diff_eq}), one gets $E+B={\rm const}\equiv -1$, 
for which $F=\frac 12\D U$. Letting $H\equiv -(2E+1)^{-1}$, the metric can be cast into 
the Gibbons-Hawking space~\cite{Gibbons:1979zt}
\begin{align}
\label{}
\D s^2=H^{-1}(\D t+\omega)^2+H\D {\bf x}^2 \,, 
\end{align}
(\ref{domega}) is simplified to $\vec \nabla\times \vec \omega =\vec \nabla H$ and 
the Maxwell equation now reads $\vec \nabla^2 H=0$. 
The Killing spinor is 
\begin{align}
\label{KS_GH}
\epsilon=  \frac 12\left(1-i \gamma_1 H^{-1/2}\right) 
(1+\gamma_5) \epsilon_0 \,. 
\end{align}
Following the same argument of \cite{Klemm:2015mga}, 
the self-duality of the Riemann tensor follows from the Killing spinor equation. 
Note also that the chiral part of (\ref{KS_GH}), $\zeta_+=\frac 12(1+\gamma_5)\epsilon_0$,
is a covariantly constant spinor $\nabla_\mu \zeta_+=0$.

\section{Petrov-D LeBrun metric}
\label{sec:typeD_LeBrun}

We have shown in section~\ref{sec:SUSY} that the self-dual PD metric can be cast into the Przanowski-Tod form, which is conformal to the LeBrun metric. This appendix is devoted to this LeBrun metric, which is obtainable by the self-dual limit of ambi-K\"ahler metric 
(\ref{Kahler_conformalCarter}) and  is given by (we will omit the `hat' in what follows)
\begin{align}
\label{D_LeBrun}
\D s^2=\frac{1}{(q+p)^2} \left\{
\frac{Q(q)}{q^2-p^2} (\D \tau-p^2 \D \sigma)^2 +(q^2-p^2)\left(\frac{\D q^2}{Q(q)}-\frac{\D p^2}{P(p)}\right)+\frac{P(p)}{p^2-q^2} (\D \tau-q^2 \D \sigma)^2 
\right\}\,,
\end{align}
for which the structure functions are given by
\begin{align}
\label{}
Q(q)=a^2-m^2+2m q-q^2+2bmq^3+q^4 \left(b^2(a^2-m^2)-\frac{\Lambda}3\right) \,, \qquad 
P(p)=Q(-p)\,. 
\end{align}
This metric solves the Einstein-Maxwell system without a cosmological constant (\ref{LeBrun_Ein}), where
the Maxwell field $\ma F=\D \ma A$ is given by 
\begin{align}
\label{LeBrun_gauge}
\ma A= \frac{(p+q)^3}{4\sqrt 2(p-q)}&\left\{-\left[
\partial_q\{(p+q)^{-4}Q(q)\}+\frac{4q}{(p+q)^4}
\right](\D \tau -p^2\D \sigma) \right. \notag \\
& \left. +\left[
\partial_p\{(p+q)^{-4}P(p)\}+\frac{4p}{(p+q)^4}
\right](\D \tau -q^2\D \sigma)\right\}\,. 
\end{align}
One can similarly compute quantities in the ($u, v$) coordinates, but we shall not pursue this line here. 

Since the conformal transformation does not change the 
algebraic type of curvatures, the resulting LeBrun metric (\ref{D_LeBrun}) also belongs to type D (and self-dual).  
Namely, the only nonvanishing Weyl scalar in the dyad notation is~\cite{Goldblatt:1994rx}
\begin{align}
\label{Psi2}
\Psi_{0011}^+\equiv -C_{\mu\nu\rho\sigma}l^\mu m^\nu \bar m^\rho \bar l^\sigma =
-\frac{2m(p+q)^2(1+bpq)}{(p-q)^3} \,, 
\end{align}
where have taken  the complex basis as 
\begin{align}
\label{lmbasis}
l=\frac 1{\sqrt 2}(e^1+i e^2) \,, \qquad m=\frac 1{\sqrt 2}(e^3+i e^4) \,, 
\end{align}
with 
\begin{align}
e^1& =\sqrt{\frac{Q(q)}{q^2-p^2}}
\frac{(\D \tau -p^2 \D \sigma )}{p+q}
\,, \qquad e^2 = \sqrt{\frac{q^2- p^2 }{Q(q)}} \frac{\D q}{p+q}
\,, \nonumber \\
e^3&= \sqrt{\frac{p^2-q^2}{P(p)}}\frac{\D p}{p+q} \,, \qquad 
e^4= \sqrt{\frac{P(p)}{p^2-q^2}}\frac{(\D \tau -q^2 \D
 \sigma )}{p+q} \,, 
\end{align}
The complex null vectors (\ref{lmbasis}) are geodesic and shear-free 
\begin{align}
\label{}
\kappa\equiv m^\mu l^\nu \nabla_\nu l_\mu =0\,, \quad 
\sigma\equiv l^\mu m^\nu \nabla_\nu m_\mu =0\,, \quad 
\kappa'\equiv  l^\mu  l^\nu \nabla_\nu \bar m_\mu =0\,, \quad 
\sigma'\equiv  l^\mu \bar m^\nu \nabla_\nu \bar m_\mu =0\,, \quad 
\end{align}
(\ref{Psi2}) implies that the metric with $m=0$ is conformally flat. 

One may thus wonder that the type-D LeBrun metric is also isometric  to the 
self-dual PD metric with $\Lambda=0$, since it belongs to Petrov-D and solves the Einstein-Maxwell field equations. However, this is not the case. A principal reason is that the Maxwell field cannot vanish, 
since the anti-self-dual part of the Maxwell field $\ma F$ is given by the K\"ahler form [see (\ref{LeBrun_Max})],  whereas the self-dual PD metric with $\Lambda =0$ reduces to the Euclid space for $m=0$. 
Another important fact for this observation is that in the above basis 
the `radiative part' of Maxwell scalars is also nonvanishing
\begin{align}
\label{}
\Phi^+_{11} \equiv \ma F_{\mu\nu}l^\mu m^\nu =-\frac{\sqrt{-P(p)Q(q)}}{\sqrt 2 (p^2-q^2)} \,.
\end{align}
This is in sharp contrast with the PD family, for which the nonvanishing Maxwell scalars are only the 
`Coulomb part' $\Phi^\pm _{01}=-\frac 12 F_{\mu\nu}(l^\mu\bar l^\nu \pm m^\mu\bar m^\nu)$. 
To the best of our knowledge, this is the first instance of Petrov-D space(time)s in which the non-null  Maxwell two-form is not aligned.

\end{document}